\documentclass[%
 aip,
 amsmath,amssymb,
 reprint,%
]{revtex4-1}

\usepackage{graphicx}% Include figure files
\usepackage{dcolumn}% Align table columns on decimal point
\usepackage{bm}% bold math
\usepackage{amssymb}
\usepackage{amsmath}
\usepackage{amsfonts}
\usepackage{mathrsfs}
\usepackage{color}
\usepackage{ulem}
\usepackage{physics}
\usepackage[english]{babel}
\usepackage[utf8]{inputenc}
\usepackage[T1]{fontenc}
\usepackage{mathptmx}
\usepackage{etoolbox}

\makeatletter
\def\@email#1#2{%
 \endgroup
 \patchcmd{\titleblock@produce}
  {\frontmatter@RRAPformat}
  {\frontmatter@RRAPformat{\produce@RRAP{*#1\href{mailto:#2}{#2}}}\frontmatter@RRAPformat}
  {}{}
}%
\makeatother
\begin{document}

\preprint{AIP/123-QED}
\newcommand{\half}{\frac{1}{2}}
\newcommand{\pcl}[1]{#1_{\mathrm{p}}}
\title[A mechanical analog of Bohr's atom based on de Broglie's double-solution approach ]{A mechanical analog of Bohr's atom based on de Broglie's double-solution approach }
% Force line breaks with \\
\author{P. Jamet}
\email{pierre.jamet@neel.cnrs.fr} 
 \affiliation{Univ. Grenoble Alpes, CNRS, Grenoble INP, Institut Neel, F-38000 Grenoble, France}
\author{A. Drezet}%
 \email{aurelien.drezet@neel.cnrs.fr}
\affiliation{Univ. Grenoble Alpes, CNRS, Grenoble INP, Institut Neel, F-38000 Grenoble, France}

\date{\today}% It is always \today, today,

\begin{abstract}
Motivated by recent developments of hydrodynamical quantum mechanical analogs [J. W. M. Bush, Annu. Rev. Fluid Mech. 47, 269–292 (2015)] we provide a relativistic model for a classical particle coupled to a scalar wave-field through a holonomic constraint. In presence of an external Coulomb field we define a regime where the particle is guided by the wave in a way similar to the old de Broglie phase-wave proposal. Moreover, this dualistic mechanical analog of the quantum theory is reminiscent of the double-solution approach suggested by de Broglie in 1927 and is able to reproduce the Bohr-Sommerfeld semiclassical quantization formula for an electron moving in a atom.   
\end{abstract}

\maketitle

\begin{quotation}
The old atomic model proposed by Bohr for a single electron orbiting in a Coulombian potential constitutes a paradigmatic example of paradoxical physics conflicting with classical intuitions. Here, modifying a old proposal made by de Broglie to explain orbital quantization using a pilot-wave guiding the particle, we develop a realistic model to make sense of Bohr's theory. Our approach considers a classical particle nonlinearly coupled to a scalar wave-field through a holonomic constraint. In presence of an external Coulomb field we define a regime where the particle is guided by the wave and reproduces the well known Bohr-Sommerfeld quantization rule for circular orbits.    
\end{quotation}

\section{Introduction}
\indent  One of the most remarkable feature of quantum mechanics is the prediction of stable electronic motions in atoms. Nowadays, and as every physics students knows, atomic orbitals are easily computed by solving the stationary Schr\"odinger's equation in a central Coulomb potential~\cite{Schrodinger1926}. However, in the early times of the quantum era  Bohr (following results by Planck, Einstein, Nicholson and Sommerfeld) already obtained a semiclassical model of an electron circular orbit in the Hydrogen atom~\cite{Bohr1918}. The method of Bohr, which was subsequently generalized by Sommerfeld to elliptical and relativistic motions~\cite{Sommerfeld}, starts with the quantization of the action variable $J=\oint P dq=2\pi n\hbar$ (with $P$ and $q$ two conjugate canonical variables, $n\in \mathbb{N}$ and $\hbar$ the reduced Planck constant) calculated along a periodic motion. However, despite tremendous success and theoretical agreements in reproducing spectrometry experiments with various atoms the Bohr-Sommerfeld model seated on an unclear basis mixing elements of classical mechanics with unjustified quantization rules.  As we know, progress in the physical understanding started when de Broglie added a circulatory wave propagation going along with the particle orbital motion~\cite{debroglie1923,debroglie1925} (the idea to introduce oscillations of an electric or mechanical medium to explain atomic spectra was originally proposed  by Nicholson and Brillouin~\cite{Nicholson,Brillouin}). In this approach, the momentum $P$ of the particle is associated with the wave vector $k$ of the wave through the formula $P=\hbar k$ and the quantization  $J=2\pi n$ becomes a stationary condition for the phase  $\varphi$ of the wave around the closed orbit: $\Delta \varphi=\oint k dq=2\pi n$. De Broglie was strongly motivated by the analogy  existing between Fermat's principle  in optics  and Maupertuis's least-action  theorem in classical mechanics (an analogy already exploited by Hamilton and Jacobi).  Generally, the textbook historical explanations concerning de Broglie’s method stop here and the mathematical development goes then with the more precise Schr\"odinger equation propagating in the configuration space and thereby abandoning the dualistic association of a wave with a particle trajectory advocated by de Broglie. Following the work of Born~\cite{Born}, the wave ultimately becomes probabilistic and the motion of the particle in space-time completely disappears from the quantum formalism.  Nevertheless, de Broglie didn't agree with theses developments and tried to obtain a more physical and deterministic interpretation of the wave mechanics in which particles and waves move together in space-time. In particular,  he and later Bohm developed a `pilot-wave interpretation' of quantum mechanics~\cite{Valentini,Bohm1952,Hiley} which is empirically equivalent to the standard quantum approach.\\
\indent In the last decades interest for the pioneering mechanical modeling of de Broglie resurrected  with the development of new fluid mechanical analogs of quantum mechanics by Couder and coworkers (for reviews see \cite{Bush2015,Bush2015b,Bush2021,BushChaos}) in which a particle droplet bouncing on a vibrating oil bath reproduces several paradigmatic features of quantum mechanics such as the wave-particle duality in double slit experiments~\cite{Couder2006}, the quantum tunneling effect~\cite{Couder2009,Bush2017} and most importantly for us: Stationary quantized states in external potentials~\cite{Fort2010,Harris2013,Gilet2016,Shinbrot2019}. These works are remarkable by their analogy with an early proposal by de Broglie named `double solution' \cite{debroglie1927,debroglie1956} in which a wave field  $u(t,\mathbf{x})$ propagating in the usual 4D space-time guides a field singularity acting as a particle and synchronized with the $u-$wave during its motion(see also~\cite{Dagan2020,Durey2021} for recent quantum hydrodynamical models similar to de Broglie's proposal).\\
\indent In a previous article~\cite{Drezet2020}, based on early proposals by Boudaoud \textit{et al.} \cite{Boudaoud1999,Borghesi2017}, we developed a mechanical analog of the double solution proposal by using the  motion of a sliding bead, i.e., a `particle',  on a vibrating string to model wave-particle duality. In this approach the transverse wave $u(t,x)$ propagating along the  $x$ direction carries the particle motion $x_p(t)$ along the same direction as the wave. A phase matching condition between the sliding particle and the wave is leading to a quantum-like guidance  of the particle by the wave in a way reminiscent of the so-called de Broglie-Bohm quantum interpretation~\cite{Hiley}. This first model ideally reproduces some features of the phase-wave introduced by de Broglie in 1923~\cite{debroglie1923,debroglie1925} at least for the simple case corresponding to the linear and uniform motion as it was first considered by de Broglie.\\ 
\indent  In the present work we extend the previous 1D model and propose a more sophisticated approach with a complex wave field $u(t,\mathbf{x})\in \mathbb{C}$ propagating in the 3 dimensional space. The model takes into account the presence of an external Coulomb field acting on the charged particle with trajectory $\mathbf{x}_p(t)$. This proposal is built in a fully covariant and relativistic framework which agrees with the original methodology proposed by de Broglie~\cite{debroglie1923,debroglie1925}. In turn, this allows us to study the quantized circular motion of a relativistic particle  in the Coulomb field and allows us to reproduce the well-known Bohr-Sommerfeld quantization formula for this Hydrogen-like atom.  We stress that, like in de Broglie's double solution~\cite{debroglie1927,debroglie1956}, our model reveals the fundamental role played by a group and  phase  contributions in the $u$-wave. In turn, this decomposition imposes strong constraints on the physical properties of the particle, i.e., its orbital motion but also its mass and electric charge.\\
\indent The layout of this work is as follow:  In Sec.~\ref{sec:2} after a short reminder concerning de Broglie original idea about phase-waves we describe our relativistic model starting from a covariant Lagrangian formulation. In Sec.~\ref{sec:3} we discuss the solution of our systems of equations  leading   to an entangled dynamics between particle and wave  (the particle being guided  by the wave).  We show how to recover the Bohr-Sommerfeld quantization  approach, i.e., at least for circular orbits, and subsequently  discuss  the   constraints and limitations of our model in Sec.~\ref{sec:4}.  

%%%%%%%ù

\section{The relativistic atom}\label{sec:2}

\subsection{The historical de Broglie’s derivation of the Bohr-Sommerfeld quantization condition}\label{sec:2a}

\indent In 1923-1924, Louis de Broglie proposed a model~\cite{debroglie1923} in order to reproduce the circular uniform motion of an electron in an atom, which in turn explained  the famous Bohr quantization hypothesis~\cite{Bohr1918}:
\begin{equation}
L_z = n\hbar
\end{equation}
with $n \in \mathbb{N}$ and $L_z = m_e v_e r_e$ the orbital angular momentum of the electron for an orbit of radius $r_e$ and velocity $v_e$.\\
\indent To understand de Broglie's insight let us consider an electron of mass $m_e$ orbiting with constant velocity $v_e$ around a nucleus. De Broglie’s idea, centered around the notion of wave-particle duality, was first to match the relativistic rest energy of the electron $m_e c^2$ with an oscillatory energy $\hbar \omega_e$, i.e., a  local `clock', so that the electron undergoes an internal motion of the form
\begin{equation}
\mathrm{e}^{-i\omega_e \tau}=\mathrm{e}^{-i\omega_e^\prime t}
\end{equation}
with $\omega_e^\prime = \omega_e\sqrt{1-v_e^2/c^2}$ the frequency of the electron as seen by an external observer in the laboratory reference frame (here $\tau$ is the proper time associated with the internal clock and  $t=\frac{\tau}{\sqrt{1-\frac{v_e^2}{c^2}}}$ is the relativistically dilated time interval in the laboratory frame where the clock is moving at a constant speed $v$). 
Then, he introduced a phase wave on the electron’s path
\begin{equation}
\mathrm{e}^{-i\omega \left(t -  x \frac{v_e}{c^2}\right)}.
\end{equation}with  $x=r_e\varphi$ a coordinate along the circular orbit ($\varphi$ is the azimuthal angle in the plane of the orbit).
In vacuum this phase wave has a velocity $v_\varphi = c^2/v_e > v_e$, so that it will catch up with the electron after a time $\delta t$:
\begin{equation}
v_\varphi\delta t = L + v_e\delta t\label{eq:vitessephase}
\end{equation}
with $L$ the length of the orbit. We find that this time $\delta t$ is
\begin{equation}
\delta t = \frac{L}{v_\varphi - v_e} = \frac{v_e}{c^2}\frac{L}{1 - \frac{v_e^2}{c^2}}.
\end{equation}
De Broglie introduced his famous `phase-harmony‘ condition telling that the phase wave and the internal oscillation of the electron must be locked in phase in order to have a stable motion. On top of that, these two phases must be multiples of $2\pi$, since we are on a circular path and we have to impose a periodicity condition:
\begin{equation}
\omega_e^\prime\delta t = \frac{m_e c^2}{\hbar} \frac{1}{\sqrt{1 - \frac{v_e^2}{c^2}}}\frac{v_e}{c^2}L = 2\pi n \label{eq:debrogliephasematching}
\end{equation}
Now given that $L = 2\pi r_e$, and $P_e=m_e v_e /\sqrt{1-v_e^2/c^2}$ the relativistic linear momentum, we finally get:
\begin{equation}
2\pi P_e r_e =\oint P_edx= 2\pi L_z = n h \label{eq:debrogliephasematchingBis}
\end{equation}
Thus we get back Bohr's quantization condition.\\
\indent This derivation is sketchy since de Broglie originally assumed no external potential acting on the particle. The model actually corresponds to the case of a particle constrained to move along a closed loop.  In a more realistic case the particle is moving along a circular orbit in a central potential $U(r)$. The phase/action velocity is given by 
\begin{eqnarray}
v_\varphi=\frac{E_e}{P_e} =\frac{c^2}{v_e}\frac{E_e}{E_e-U}
\end{eqnarray} where 
\begin{eqnarray}
P_e=\frac{m_e v_e}{\sqrt{1-\frac{v_e^2}{c^2}}},  E_e=\frac{m_ec^2}{\sqrt{1-\frac{v_e^2}{c^2}}} +U(r)
\end{eqnarray} are the particle linear momentum and total energy respectively.  The particle velocity is given by Hamilton's equation
\begin{eqnarray}
v_e=\frac{\partial E_e}{\partial P_e}=c^2\frac{P_e}{E_e-U}
\end{eqnarray} which is subsequently identified with Rayleigh's group velocity $v_g=\frac{\partial \omega}{\partial k}$ using the iconic quantum relations $P_e=\hbar k$, $E_e=\hbar \omega$. Moreover,  the phase matching condition Eq.~\ref{eq:vitessephase}  still holds and instead of 
Eqs.~\ref{eq:debrogliephasematching},\ref{eq:debrogliephasematchingBis}   we get:
\begin{eqnarray}
-\mathcal{L}_e\delta t=P_eL=\oint P_edx= 2\pi L_z = n h
\end{eqnarray} which again recovers Bohr's quantization condition. In this formula $\mathcal{L}_e=P_ev_e-E_e=-m_e\sqrt{1-v_e^2/c^2}-U$ is the Lagrangian of the particle and $-\mathcal{L}_e/\hbar$ plays the role of the clock frequency $\omega_e^\prime = \omega_e\sqrt{1-v_e^2/c^2}$ used in the original de Broglie deduction.\\
\indent This idea of a moving clock synchronized with a guiding wave is the hallmark of de Broglie's conception of quantum mechanics. If we generally write $S_e(t)=\int_{(C)}^t dt' \mathcal{L}_e(t')$ the action integral along the trajectory $C$ followed by the particle the phase harmony condition of de Broglie  reads   
\begin{eqnarray}
\frac{d}{dt}S_e=\mathcal{L}_e=\hbar\frac{d}{dt}\varphi
\end{eqnarray} 
where $\hbar\frac{d}{dt}\varphi:=-\hbar \omega_e^\prime(t)=-\hbar \omega_e(t)\sqrt{1-v_e^2/c^2}$ generally defines a time dependent frequency $\omega_e(t)$ along the path. \\
\indent In the following years after his PhD thesis~\cite{debroglie1925}, de Broglie tried to make sense of his phase-wave hypothesis. As the name suggests, this generally superluminal wave is not strictly a physical object and does not carry any energy, but is rather a clever way to make sense of quantum features such as wave-particle duality and quantization conditions. His goal was to build a more realistic mechanical model using a `physical' wave $u(t,\mathbf{x})$ that one could interpret as being the particle. He called `double solution' his new proposal~\cite{debroglie1927} in which a wave-field $u$ carrying the particle energy coexisted with the more conventional $\psi-$field used in quantum  (wave) mechanics.   In the later sections of this paper, we propose a physical model for this  de Broglie wave using a complex scalar field $u(t,\mathbf{x})$ which we can decompose as a group-wave carrying the energy (i.e., the particle), and a phase-wave which is actually predating the $\psi-$wave solution of Schr\"odinger's equation.
%%%%%%
\subsection{A relativistic atomic model coupling a wave-field and a particle }\label{sec:2b}
\indent In this section, we will derive the equations of motion of our model using natural units ($c = \hbar = 1$) for simplicity, and reintroduce the physical constants when necessary. Furthermore, we consider the Minkowski metric $\eta_{\mu,\nu}$ with signature $(1,-1,-1,-1)$ in the following.
We start with the relativistic action
\begin{equation}
    \begin{split}
		I = &-\int\bqty{\pcl{m} - \half m_{\mathrm{p}}\sigma\pqty{\vqty{\dot{z}(\tau)}^2 - \pcl{\Omega}^2 \vqty{z(\tau)}^2}}\dd\tau \\
		&+ \int \left\lbrace \mathcal{N}(\tau)\bqty{z(\tau) - u(\pcl{x}(\tau))}^*\right. \\&+ \left. \mathcal{N}^*(\tau)\bqty{z(\tau) - u(\pcl{x}(\tau))}\right\rbrace\dd\tau \\ 
		&- e\int A(\pcl{x}(\tau))\pcl{\dot{x}}(\tau)\dd\tau + T\int(Du)(Du)^*\dd^4 x 
	\end{split}
	\label{eq:action}
\end{equation}
Let us specify what each of these terms represent. The first term 
\begin{eqnarray}
-\int\bqty{\pcl{m} - \half m_{\mathrm{p}}\sigma\pqty{\vqty{\dot{z}(\tau)}^2 - \pcl{\Omega}^2 \vqty{z(\tau)}^2}}\dd\tau \label{relatos}
\end{eqnarray} (where $\dd\tau=\sqrt{\eta_{\mu,\nu}dx^\mu dx^\nu}$ is a proper time interval along the particle trajectory $x_p^\mu:=[t,\mathbf{x}_p]\in \mathbb{R}^4$ and where $\frac{d}{d\tau} z(\tau):=\dot{z}(\tau)$) is a modification of the relativistic action for a free particle of mass $\pcl{m}$, to which we added an internal oscillating degree of freedom $z(\tau) \in \mathbb{C}$, with  $\pcl{\Omega}$ the typical pulsation of this harmonic oscillator. This leads us to interpret $\pcl{m} - \half m_{\mathrm{p}}\sigma\pqty{\vqty{\dot{z}(\tau)}^2 - \pcl{\Omega}^2 \vqty{z(\tau)}^2}$ as a kind of varying relativistic mass (with $\sigma $ a mass correction coefficient).  In a previous paper~\cite{Drezet2020}, we used this coordinate $z$ in a mechanical sense to represent the transverse oscillations on a string. Here however, to account for the 3-dimensional and relativistic nature of our system, we take it as an internal degree of freedom (more on this analogy is discussed below). This degree of freedom is interacting with a field $u(t,\mathbf{x})=u(x)\in \mathbb{C}$ (with $x^\mu:=[t,\mathbf{x}]\in \mathbb{R}^4$) by way of the constraint written in the second integral:
\begin{eqnarray}
\int \left\lbrace \mathcal{N}\bqty{z - u(\pcl{x})}^*+ \mathcal{N}^*\bqty{z- u(\pcl{x})}\right\rbrace\dd\tau
\end{eqnarray}
which involves two additional complex scalar fields $\mathcal{N}(\tau)$, $\mathcal{N}^\ast(\tau)$ used to model a holonomic constraint for the particle.  In the model we also include an external electromagnetic  potential $A^\mu(x):=[V(x),\mathbf{A}(x)]$, which will affect the particle with electric charge $e=-|e|<0$ according to the third integral $- e\int A(\pcl{x})\pcl{\dot{x}}\dd\tau $. Finally, we have the Lagrangian density for a scalar fields $u$, $u^\ast$, where we replaced the partial derivatives $\partial_\mu:=[\partial_t,\boldsymbol{\nabla}]$ with covariant derivatives $D_\mu = \partial_\mu + \mathrm{i} e' A_\mu(x)=[\partial_t+ie'V(x),\boldsymbol{\nabla}-ie'\mathbf{A}(x)]$, again to take into account this external potential, and keep a covariant and gauge invariant relativistic formulation. This corresponds to the description of charged electric fluid described by two fields $u,u^\ast$. We point out that for the sake of generality the electric charge $e'$ of the charged fluid  is not necessarily equal to $e$. Ultimately, its value could even vanish. We will go back to this issue in Sec.~\ref{sec:3c}  \\
\begin{table}[h]
\begin{tabular}{l||c}
 Parameters & Physical meaning \\
\hline
\hline
$u(x)$  & \textrm{Fundamental field}\\
\hline
$A^\mu(x)$  & \textrm{Electromagnetic 4-vector potential}\\
\hline
$z(\tau)$  & \textrm{Internal oscillator} \\
\hline
$\mathcal{N}(\tau))$  & \textrm{Internal reaction force acting upon the particle} \\
\hline
$\pcl{x}^\mu(\tau)$  & \textrm{4-vector position of the particle} \\
\hline
$T$  & \textrm{Tension of the field} \\
\hline
$\pcl{\Omega}$  & \textrm{Internal oscillator pulsation} \\
\hline
$m_{\mathrm{p}}$  & \textrm{Bare particle mass} \\
\hline
$\sigma$  & \textrm{Coupling constant} \\
\hline
$e$  & \textrm{Particle electric charge} \\
\hline
$e'$  & \textrm{$u-$field electric charge} \\
\end{tabular}
\caption{Table summarizing the different parameters  of the model. }\label{table1}
\end{table}
\indent We summarize the different fundamental variables  and constants of our model in Table~\ref{table1}. In order to physically interpret our model it is probably useful to compare the present work with our previous non-relativistic approach published in \cite{Drezet2020} for a 1D  string-like model. First, if we take a more mechanical analogy in 2D instead of 3D $u(t,x,y)$ could be interpreted as the vibration of an elastic membrane with tension $T$. The $u-$ field could thus be  seen as the transverse motion along the $z$ (vertical) direction of the membrane and the point on the surface is labeled by 2D coordinates $x,y$. This mechanical analogy makes sense for small transverse vibrations (i.e. like in the 1D mechanical analog  of \cite{Drezet2020}). Here, our covariant model works in a 4D space-time and the $u-$field is complex rather than real but this generalization is not mandatory and only makes the framework more elegant and symmetrical. The internal vibration $z(\tau)$ is just, if we follow this membrane analogy, the height at which the particle is located on top of the surface, i.e., $z(\tau):=\pcl{z}(t)$. This transverse motion should not be confused with the $\pcl{x}^\mu(\tau)$ particle motion which in the 2D analogy is just the set $t, \pcl{x}(t), \pcl{y}(t)$  describing the in-plane motion of the particle.   In 4D  it is more judicious to call $z(\tau)$ an internal motion acting in a different space (also the proper time label $\tau$ helps to make the theory fully covariant). The other variables have also a clear physical meaning in this analogy: For instance,  $\pcl{\Omega}$ is a mechanical pulsation associated with a vertical restoring force acting upon the particle.   Taking the non-relativistic limit of Eq.~\ref{relatos} and using $z(\tau):=\pcl{z}(t)$ we obtain 
\begin{eqnarray}
\int[-\pcl{m} + \half m_{\mathrm{p}}\dot{\pcl{\mathbf{x}}}(t)^2+\half m_{\mathrm{p}}\sigma(|\dot{\pcl{z}}(t)|^2 - \pcl{\Omega}^2 |\pcl{z}(t)|^2)]\dd t\nonumber\\ \label{relatosbis}
\end{eqnarray} 
which is indeed describing the motion of an harmonic oscillator \cite{Drezet2020}. Moreover, the presence of the electric  field  acting on the particle with charge $e$ can easily be included in this mechanical analogy. Finally, the charge   $e'$ associated with the oscillating medium is more difficult to physically interpret in the membrane analogy. Still,  the   $e'$ charge together with the complex nature of the $u-$ field allow us to introduce gauge invariance and  covariant derivative $D_\mu = \partial_\mu + \mathrm{i} e' A_\mu(x)$ in the formulation which is fine in the context of a fundamental quantum theory.   Moreover, the model developped here is so robust that even the case $e'=0$ can be used to model a Bohr atom as we will see in Sec.~\ref{sec:3c}.   In such a case the membrane mechanical analogy is perfectly valid and could be used for developing a possible experimental demonstrator.      \\ 
\indent Now going back to our model, we can see that obtaining the equations of motion is straightforward using the Euler-Lagrange equations (the full derivation of which can be found in the appendix). We choose to consider cases where there is no longer any interaction between the particle and the field, \textit{i.e.}
\begin{eqnarray}
\mathcal{N}(\tau) = 0, & \mathcal{N}^\ast(\tau) = 0, 
\end{eqnarray}  as was motivated in \cite{Drezet2020}. This regime, hereafter referred to as transparency, strongly simplifies the dynamics. The goal here is to find stable solutions for the motion of the particle, and construct a wave-particle duality model. We can ultimately turn back to investigating the chaotic regimes where the field and the particle start interacting with each other, and as such consider dynamical cases like, for example, atomic transitions and photon emissions: This will be the subject of future works. \\
\indent In this transparency regime we get the following equations:  First we obtain the condition 
\begin{equation}
	z(\tau) - u(t,\vb{x}_{\mathrm{p}}) = 0,
	\label{eq:constr}
\end{equation}
which is very general in our theory (i.e., independent of the transparency regime) and models the holonomic constraint that we impose between the field and the particle. This constraint is central in our theory since, as show below, it allows us to recover the phase harmony condition introduced by de Broglie.  \\
\indent Moreover, for $z(\tau)$ we have also the simple equation
\begin{equation}
	\ddot{z}(\tau) + \Omega_{\mathrm{p}}^2 z(\tau) = 0
\end{equation}
which gives us a relativistic harmonic motion of the form
\begin{equation}
	z(\tau) = z_0\mathrm{e}^{-\mathrm{i}\Omega_{\mathrm{p}}\tau}.\label{eq:zmotion}
\end{equation}
\indent Assuming Eq.~\ref{eq:constr} and Eq.~\ref{eq:zmotion} we deduce the equation of motion for the position of the particle
\begin{equation}
	m_{\mathrm{p}}\pqty{1 + \sigma\Omega_{\mathrm{p}}^2 |z_0|^2} \ddot{x}_{\mathrm{p}\mu}(\tau) = eF_{\mu\nu}(x_{\mathrm{p}}(\tau))\dot{x}_{\mathrm{p}}^\nu(\tau)
	\label{eq:particle_motion}
\end{equation}
where $F_{\mu\nu} = \partial_\mu A_\nu - \partial_\nu A_\mu$ is the electromagnetic tensor. This equation is completely analogous to Newton’s second principle, where the Lorentz force accelerating the particle comes directly from the applied external electromagnetic field (the standard Lorentz force involved in Eq.~\ref{eq:particle_motion} reads $\vb{F} = e(\vb{E}+\vb{v}_{\mathrm{p}}\times\vb{B})$), with the introduction of an effective and constant mass term $m_{\mathrm{eff.}}:=m_{\mathrm{p}}\pqty{1 + \sigma\Omega_{\mathrm{p}}^2 |z_0|^2}$ to account for the oscillatory motion $z(\tau)$. We mention that if we leave the transparency regime this effective mass is generally varying with time along the trajectory. Furthermore, in this general regime we have an additional contributing force in the right hand side  of Eq.~\ref{eq:particle_motion}  reading $\mathcal{N}\partial_\mu u^\ast(x_p)+\mathcal{N}^\ast\partial_\mu u(x_p)$. This force depends on the $u$-field gradient leading to corrections on the classical-like equation of motion Eq.~\ref{eq:particle_motion} (the dynamics is derived in the Appendix).\\
\indent We emphasize that in the transparency regime (where  Eq.~\ref{eq:particle_motion} holds true)  the dynamics is also derived from the effective action:
 \begin{equation}
		S_{\mathrm{eff.}} = \int[ -m_{\mathrm{eff.}}-eA(\pcl{x}(\tau))\pcl{\dot{x}}(\tau)]\dd\tau. 
	\label{eq:actioneffective}
\end{equation}
\indent Finally, for the $u-$field  we have a d'Alembert-like equation with covariant derivatives
\begin{equation}
	D^2 u(x) = 0.
	\label{eq:field_motion}
\end{equation}
It is interesting to note that if  we leave the  transparency regime $\mathcal{N}=0$ Eq.~\ref{eq:field_motion} becomes
\begin{eqnarray}
 	D^2 u(x) = -\frac{1}{T}\int d\tau \mathcal{N}(\tau)\delta^4(x - \pcl{x})\nonumber\\
 	=-\frac{\mathcal{N}_t}{T}\sqrt{1-\pcl{v}^2(t)}\delta^3(\mathbf{x} - \pcl{\mathbf{x}}(t))\label{eq:field_source}
 \end{eqnarray}
with $\mathcal{N}_t=\mathcal{N}(\tau(t))$ and $\tau(t)=\int_{(C)}^t dt'\sqrt{1-\pcl{v}^2(t')}$ the proper time at time $t$ integrated along the path $C$ followed by the particle. Eq.~\ref{eq:field_source} includes a source term inducing the radiation of scalar $u-$waves by the particle if $\mathcal{N}\neq 0$. In the present problem the transparency regime actually decouples the field from the source leading to the homogeneous Eq.~\ref{eq:field_motion} for $u$.\\
\indent  We also stress that Eq.~\ref{eq:field_motion} is very similar to the paradigmatic   Klein-Gordon equation
\begin{equation}
	D^2 u(x) +\omega_0^2 u(x)=0,
	\label{eq:field_motionKG}
\end{equation}
where $\omega_0$ is the `Compton' frequency usually acting as a mass. For the present work exploiting the continuous spectrum of the wave equation  the condition  $\omega_0=0$ is the simplest choice (as shown in  Sec.~\ref{sec:3c}). This makes sense since in our model the frequency $\omega_0$ is not identified with the particle mass but rather characterizes the $u-$field.  Moreover, for generality we mention that if we add in the action $I$ a term $-T\int\omega_0^2 uu^\ast\dd^4 x $ we obtain the field equation: 
\begin{eqnarray}
 	D^2 u(x) +\omega_0^2 u(x)=-\frac{1}{T}\int d\tau \mathcal{N}(\tau)\delta^4(x - \pcl{x}).\label{eq:field_sourceKG}
 \end{eqnarray}
%%%%%%%%%%%%%%%%%%%%%%%%%%%%
\section{Solutions of the equations of motion}\label{sec:3}
\subsection{The transparency regime and de Broglie's wave motion}\label{sec:3a}
\indent As mentioned before, what we are looking for are configurations where the particle and field evolve independently without interaction but with the holonomic constraint imposing a local phase matching between $u$ and $z$. We know from a previous work~\cite{Drezet2020} that such a transparency regime is possible when the total field is expressed as the sum of two counter-propagating waves $u = u_+ + u_-$, with each component having the right Doppler-shifted frequency.  For now, let us restrict ourselves to a uniform and circular motion on the equatorial plane, \textit{i.e.} where the velocity $\pcl{\mathbf{v}}$ remains constant in amplitude, and the only spatial degree of freedom is the azimuthal angle $\varphi$ on a circle of radius $\pcl{r}$. This development will be further justified once we look at the full 3D case, and consider the field $u(t,r,\theta,\varphi)$ function of spherical coordinates $r$, $\theta$, and $\varphi$ in  Sec.~\ref{sec:3c}.\\ 
\indent We write the waves $u_\pm$ along the orbit of radius $\pcl{r}$ as
\begin{equation}
	u_\pm(t, \vb{x}) =\half u_0\mathrm{e}^{\mathrm{i}\pqty{\pm k_\pm \pcl{r} \varphi - \omega_\pm t}},
\end{equation}
with the periodicity condition (i.e., continuity of the wave) which imposes
\begin{equation}
	k_\pm = \frac{m_\pm}{\pcl{r}}, \quad m_\pm \in \mathbb{N}.\label{eq:quantiza}
\end{equation}
This already provides quantization conditions for the wavevectors $k_\pm$.\\  
%Note also that our wave-numbers $k_\pm$ are scalar here since we only look at the $\varphi$ direction, but they will become wave-vectors when we look at the 3D model.
\indent We now consider the total field, i.e.,  the sum of $u_+$ and $u_-$, and we get after some rearrangements:
\begin{equation}
	u(t,\vb{x}) = u_0\mathrm{e}^{\mathrm{i}(k \pcl{r} \varphi - \omega t)}\cos\pqty{\frac{k_+ + k_-}{2}\pcl{r}\varphi - \frac{\omega_+ - \omega_-}{2}t},
	\label{eq:field_e_cos}
\end{equation}
with
\begin{equation}
	k = \frac{k_+ - k_-}{2}\quad \mathrm{and}\quad \omega = \frac{\omega_+ + \omega_-}{2}.
\end{equation}
If we had a free field with two plane waves, we would have the dispersion relation $k_\pm = \omega_\pm$ as used in~\cite{Drezet2020}. However, we also want to take into account the presence of the central potential from the atom, so we make the following hypothesis
\begin{equation}
	k_\pm = \omega_\pm + \varepsilon_\pm,\label{hypo}
\end{equation}
which transforms the term inside the cosine function of Eq.~\ref{eq:field_e_cos} into 
\begin{equation}
	(\omega + \varepsilon)\bqty{r\varphi - \frac{k - \eta}{\omega + \varepsilon}t},\;\;\varepsilon = \frac{\varepsilon_+ + \varepsilon_-}{2},\;\;\eta = \frac{\varepsilon_+ - \varepsilon_-}{2}.
\end{equation}
The term in front of $t$ is the group velocity $v_g$ of our total wave, and following de Broglie’s atomic model, we identify it with the particle velocity
\begin{equation}
	v_g:=\frac{k - \eta}{\omega + \varepsilon} \equiv \pcl{v}:=\frac{\pcl{P}-eA_\varphi}{\pcl{E} - eV}	\label{eq:field_e_cosB}
\end{equation}
where $\pcl{P}$, $\pcl{E}$ are  the particle linear momentum and energy respectively. Here, in the definition of the particle velocity $\pcl{v}$ we include contributions of the external field $A^\mu:=[V,\mathbf{A}]$. In this work we only consider the scalar Coulomb field $A^0:=V(r)$ and remove the magnetic vector potential $\mathbf{A}$. We only stress that we can introduce an azimuthal magnetic vector potential   $\mathbf{A}=A_\varphi(r)\hat{\boldsymbol{\varphi}}$ to describe the Zeeman effect for our `electron' motion in the atom (this will be developed in a subsequent work).
With this hypothesis we see that if we compute the wave-field  at the position $\mathbf{x}=\pcl{\mathbf{x}}(t)$ of the particle at time $t$ we obtain 
\begin{equation}
	u(t,\pcl{\mathbf{x}}(t)) = u_0\mathrm{e}^{\mathrm{i}(k \pcl{v} - \omega)t},
	\label{eq:field_e_cosBB}
\end{equation} showing that the wave amplitude  remains constant at the position of the particle and that its phase is a linear function of $t$.   
Moreover, according to the original de Broglie hypothesis we have  $\pcl{P}= k$ and $\pcl{E} = \omega$. Yet, as Eq.~\ref{eq:field_e_cosBB} suggests it appears judicious to assume more general relations:
\begin{equation}
	\pcl{P}= bk, \textrm{ and}\quad \pcl{E} =b \omega \label{eq:definition}
\end{equation}
where $b$ is a dimensionless constant.\\ 
\indent Furthermore, we impose $\eta = 0$  and $A_\varphi=0$ in Eq.~\ref{eq:field_e_cosB} and therefore
\begin{equation}
	b\varepsilon =b\varepsilon_\pm= -eV(\pcl{r})\label{eq:epsilon} 
\end{equation}
where $V(\pcl{r})=\frac{-e}{4\pi \pcl{r}}=\frac{|e|}{4\pi \pcl{r}}$ is the external electric Coulomb potential acting upon the moving particle of charge $e=-|e|$.  We will justify the consistency of Eqs.~\ref{eq:field_e_cosB}, \ref{eq:epsilon} more rigorously once we derive the dynamics of the particle in Sec.~\ref{sec:3b}, but it is important to note that we still work with a wave-particle duality in mind, which is why Eqs.~\ref{eq:field_e_cosB}, \ref{eq:epsilon} are so crucial to our model.\\
\indent Moreover, being on a circular path, we have from Eq.~\ref{eq:quantiza} quantization conditions which impose
\begin{eqnarray}
	\pcl{P}=bk = b\frac{m_+ - m_-}{2r_{\mathrm{p}}},\nonumber \\
	\pcl{E}=b\omega =b \frac{m_+ + m_-}{2r_{\mathrm{p}}} + eV. 
	\label{eq:w_vec_freq}
\end{eqnarray}
For convenience we introduce in the following the notations
\begin{eqnarray}
	n=b\frac{m_+-m_-}{2}, \quad N=b\frac{m_++m_-}{2},
	\label{eq:nota}
\end{eqnarray}
i.e., 
\begin{eqnarray}
  bm_\pm= N \pm n.
	\label{eq:notabis}
\end{eqnarray}
\indent Importantly, the first equality of Eq.~\ref{eq:w_vec_freq}
reads 
\begin{eqnarray}
2\pi\pcl{r}\pcl{P}=\oint\pcl{P}dx=2\pi n\label{quantiza}
\end{eqnarray} (with $dx=\pcl{r}d\varphi$) which is clearly reminiscent of the Bohr-Sommerfeld quantization formula Eq.~\ref{eq:debrogliephasematchingBis} as described by de Broglie.\\ 
\indent Moreover, we now see the importance of the $b$ constant in Eq.~\ref{eq:nota}.  Indeed, if we select $b=1$ as suggested by de Broglie we obtain $n=0,\pm 1/2,\pm 1,\pm 3/2,...\in \mathbb{Z}/2$ which  introduces half-integer numbers in addition to integers usually considered in the Bohr-Sommerfeld semi-classical quantum theory. This might give rise to richer dynamics than those considered historically and one could for example speculate that this is related to the existence of an half-integer spin in quantum mechanics (this issue was already discussed by Sommerfeld and Heisenberg in the old quantum theory~\cite{Sommerfeld,Mehra}). Moreover,  observe that if we select the condition $b=2$ we recover exactly the Bohr-Sommerfeld theory. Therefore, we obtain $n=0,\pm 1,\pm 2...\in \mathbb{Z}$ as it should be. As we will show in Sec.~\ref{sec:3b} the choice of the $b$ parameter is further constrained by the value given to the electric charge $e=\sqrt{(4\pi \alpha)}$, i.e.,  as deduced from the particle dynamics.
%%%%%%%%%%%%%%%%%%%
\subsection{The quantized particle dynamics}\label{sec:3b}
\indent We are now interested in solutions of Eq.~\ref{eq:particle_motion} for circular motions of the particle. First, the term $m_{\mathrm{p}}(1 + \sigma\Omega_{\mathrm{p}}^2\vqty{z_0}^2)$ is constant for a given field and behaves as an effective mass $m_{\mathrm{eff.}}$. We also remove most of the components of the electromagnetic tensor $F_{\mu\nu}$ since we only have a static scalar potential $A^0(x) = V(\vb{x})=\frac{|e|}{4\pi r}$, which gives us in the end 
\begin{equation}
	m_{\mathrm{eff.}}\ddot{\vb{x}}_{\mathrm{p}} = -e\grad{V(\vb{x}_{\mathrm{p}})} = -\frac{\alpha}{\pcl{r}^2}\hat{\vb{r}}_{\mathrm{p}},
\end{equation} where we introduced the Sommerfeld fine-structure constant $\alpha=\frac{e^2}{4\pi}$. 
This equation also reads
\begin{equation}
	\dv{t}\pqty{m_{\mathrm{eff.}}\gamma\vb{v_{\mathrm{p}}}} = -m_{\mathrm{eff.}}\gamma\frac{v_{\mathrm{p}}^2}{\pcl{r}}\hat{\vb{r}}_{\mathrm{p}} = -\frac{\alpha}{\pcl{r}^2}\hat{\vb{r}}_{\mathrm{p}},
\end{equation}
or equivalently
\begin{equation}
	\gamma m_{\mathrm{eff.}}v_{\mathrm{p}}^2 \pcl{r} = \alpha,
\end{equation}
with $\gamma = 1/\sqrt{1 - v_{\mathrm{p}}^2}$ the Lorentz boost factor.
Furthermore, in a central potential the orbital angular momentum is a constant of motion and we have 
\begin{equation}
	J = \oint \pcl{P}\dd{x} %= \oint \gamma m_{\mathrm{eff.}}\pcl{v} \pcl{r}\dd{\varphi} = 2\pi \pcl{r}\gamma m_{\mathrm{eff.}} \pcl{v} = 
	= 2\pi n,
\end{equation}
and given that $\pcl{\mathbf{P}} =\boldsymbol{\nabla}S_{\mathrm{eff.}} = \gamma m_{\mathrm{eff.}}\mathbf{v}_{\mathrm{p}}$ (with $S_{\mathrm{eff.}} $ given by Eq.~\ref{eq:actioneffective}) this leads to 
\begin{equation}
	\gamma m_{\mathrm{eff.}}v_{\mathrm{p}} r = n.
\end{equation}
Here the angular orbital momentum $n:=L_z$ is a parameter that characterizes the orbits, our hypothesis being that the orbits are indeed quantized to recover the Bohr-Sommerfeld quantization rule and therefore $n$ is expected to be an integer. Without lack of generality we will from now consider $n\geq 0$ assuming an anticlockwise motion of the particle along the orbit. With this constraint we have naturally $m_+\geq m_- \geq 0 $ in Eq.~\ref{eq:nota}.\\
\indent In our model this quantization condition must actually match Eq.~\ref{quantiza}.  Having this in mind, we will add an index $n$ to all our orbital quantities, while also removing most of the $\mathrm{p}$ indices so as not to clutter the equations.\\
\indent In the end, we can write the velocity
\begin{equation}
	v_n = \frac{\alpha}{n}=\frac{n}{N},
	\label{eq:p_velocity}
\end{equation} leading to the momentum 
\begin{equation}
	P_n = m_{\mathrm{eff.}}\frac{\frac{\alpha}{n}}{\sqrt{1 - \frac{\alpha^2}{n^2}}}.
\label{eq:p_velocityB}
\end{equation} 
Similarly, we deduce the radius  of the orbit 
\begin{equation}
	r_n = n^2 a_0 \sqrt{1 - \frac{\alpha^2}{n^2}},
	\label{eq:p_radius}
\end{equation}
with $a_0 = 1/(m_{\mathrm{eff.}}\alpha)$ a typical distance, which is identified to an `effective' Bohr radius. Finally, we obtain the energy of our atomic system
\begin{equation}
	E_n =-\partial_t S_{\mathrm{eff.}} =\frac{m_{\mathrm{eff.}}}{\sqrt{1 - \frac{\alpha^2}{n^2}}} - \frac{\alpha}{r_n}=m_{\mathrm{eff.}}\sqrt{1 - \frac{\alpha^2}{n^2}}.
	\label{eq:p_energy}
\end{equation} All the previous formulas  require  $\alpha<1$ for consistency.  Of course, in the non relativistic limit Eq.~\ref{eq:p_energy} reduces to $E_n\simeq m_{\mathrm{eff.}} -\frac{m_{\mathrm{eff.}}\alpha^2}{2n^2}$ which is the famous Bohr quantization energy spectrum.\\  
By combining Eqs.~\ref{eq:p_velocityB} and \ref{eq:p_energy} we obtain
\begin{equation}
	v_n = \frac{P_n}{E_n+\frac{\alpha}{r_n}},
	\label{eq:p_velocityC}
\end{equation} which is identical to Eq.~\ref{eq:field_e_cosB} and justifies our identification  $v_g=v_n$ in  Sec.~\ref{sec:3a}.
Alternatively, from Eq.~\ref{eq:w_vec_freq} we get
\begin{eqnarray}
	P_n=bk = \frac{n}{r_n},\nonumber \\
	E_n=b\omega =\frac{N-\alpha}{r_n}. 
	\label{eq:w_vec_freqbis}
\end{eqnarray}
Inversely, we can express $\omega_\pm$ as functions of physical parameters associated with the particle:
\begin{eqnarray}
     b\omega_\pm=\frac{m_{\mathrm{eff.}}}{\sqrt{1 - \frac{\alpha^2}{n^2}}} (1\pm\frac{\alpha}{n}-\frac{\alpha^2}{n^2})
	\label{eq:w_vec_freqtri}
\end{eqnarray}  Note, that from Eq.~\ref{eq:w_vec_freqtri} and assuming $\omega_\pm\geq 0$  we deduce the constraint $\alpha<(\sqrt{5}-1)/2$ that is obviously satisfied with $\alpha\simeq 1/137$. \\
\indent All these results are particularly interesting, since we recover the known semi-classical formulas, with an added relativistic correction in the term $\sqrt{1 - \alpha^2/n^2}$, the same correction that appears in Sommerfeld's extension of Bohr's atomic model for circular orbits.\\
\indent Before moving to the rigorous solutions for the $u$-field, we have two more important equations of motion for the particle to derive. Let us write the Lagrangian for the particle
\begin{equation}
	\mathcal{L}_n = -m_{\mathrm{eff.}}\sqrt{1 - \frac{\alpha^2}{n^2}} + \frac{\alpha}{r_n}.
\end{equation}
Using a Legendre transform this also reads
\begin{equation}
	\mathcal{L}_nt= (P_n v_n - E_n) t = b(kv_n-\omega)t.
\end{equation} 
In the end, if we inject the Lagrangian $\mathcal{L}_n$ into the $u-$field given by Eq.~\ref{eq:field_e_cosBB} and use the constraint given by Eq.~\ref{eq:constr} we get de Broglie's phase harmony condition
\begin{equation}
	z_0\mathrm{e}^{-\mathrm{i}\Omega_{\mathrm{p}}\sqrt{1 - \frac{\alpha^2}{n^2}}t} = u_0\mathrm{e}^{\mathrm{i}\frac{\mathcal{L}_n t}{b}}.\label{eq:debrogliephasematch}
\end{equation}
This condition implies $z_0=u_0$ which is linking the amplitude of the internal vibration with the one of the wave. From Eq.~\ref{eq:debrogliephasematch} we also deduce 
\begin{equation}
	b\Omega_{\mathrm{p}} = m_{\mathrm{eff.}}\pqty{1 - \frac{\alpha^2}{n^2 - \alpha^2}}\label{mass}
\end{equation} which shows that in the present model the effective mass is quantized and depends on the quantum number $n$. More precisely, if we consider the fundamental constants of the model $\Omega_{\mathrm{p}}$ and $\sigma$ Eq.~\ref{mass} equivalently reads    
\begin{equation}
	\vqty{z_0}^2 = \frac{1}{m_{\mathrm{p}}\sigma\Omega_{\mathrm{p}}^2}\pqty{\frac{b\Omega_{\mathrm{p}}}{1-\frac{\alpha^2}{n^2-\alpha^2}} - m_{\mathrm{p}}}.\label{amplitude}
\end{equation}
which fixes the amplitude $\vqty{z_0}$ and shows that its value is quantized.  In particular, if we consider the limit $n\rightarrow +\infty$ we obtain $b\Omega_{\mathrm{p}} \simeq m_{\mathrm{eff.}}$ and $\vqty{z_0}^2\simeq \frac{1}{m_{\mathrm{p}}\sigma\Omega_{\mathrm{p}}^2}\pqty{b\Omega_{\mathrm{p}} - m_{\mathrm{p}}}$. \\
\indent One last interesting result can be obtained by combining Eqs.~\ref{eq:w_vec_freq} and \ref{eq:p_energy}, to extract a value for $\alpha$. We quickly find that
\begin{equation}
	\alpha = \frac{n^2}{N} = \frac{b}{2}\frac{(m_+ - m_-)^2}{m_+ + m_-},\label{eq:charge}
\end{equation}
or in other terms, that the electric charge $e = \sqrt{4\pi\alpha}$ is also quantized. A different way for obtaining this result is to start with the relativistic mechanical dispersion relation for a charged particle in a Coulomb potential:
\begin{eqnarray}
(E_n+\frac{\alpha}{r_n})^2-P_n^2=m_{\mathrm{eff.}}^2
\end{eqnarray}
After substitution  of Eq~\ref{eq:w_vec_freq} we obtain $N^2-n^2=m_{\mathrm{eff.}}^2r_n^2$ and finally using the formula Eq.~\ref{eq:p_radius} for  $r_n$ we recover Eq.~\ref{eq:charge} (an alternative way to deduce this relation is to apply the identity $v_n=\alpha/n=P_n/(E_n-eV)=n/N$).
\\ 
\indent Remarkably, Eq.~\ref{eq:charge} is the only place where the quantum number $N=b\frac{m_++m_-}{2}$ appears explicitly in the dynamics.   This relation is particularly interesting since it reveals the presence of two physical scales: The first one defined  by $n$ is associated with the quantum motion  of the particle in the Coulomb potential, i.e., low-energy physics, and leads to the Bohr-Sommerfeld quantization.  The second scale $N\gg n$, defines a electromagnetic scale associated with the point-like particle: It is therefore characteristic of high-energy physics.\\
\indent  It is easy to see that if $N=137$ and $n=1$ we already obtain $\alpha=1/137$ which is close to the experimental value of the fine structure constant. Since $\alpha$ is a fundamental constant Eq.~\ref{eq:charge} can actually be used to further constrain the possibilities for the value of $n$ and $N$ and therefore for the values taken by $m_\pm$. More precisely, by using Eq.~\ref{eq:charge} and Eq.~\ref{eq:notabis} we deduce:
\begin{eqnarray}
  bm_\pm= \frac{n^2}{\alpha} \pm n.
	\label{eq:notatri}
\end{eqnarray}   
This stringent  condition requiring $m_\pm\in \mathbb{N}$ is in general impossible to fulfill rigorously  for arbitrary $\alpha$ and $b$.  Moreover, observe that for $\alpha^{-1}\in\mathbb{N}$, and $b=1$ it works for any $n\in \mathbb{N}$. Furthermore, in the same conditions Eq.~\ref{eq:notatri} doesn't hold true for half integer quantum numbers $ n=1/2,3/2... \in \frac{\mathbb{N}}{2}$.   Therefore, with $b=1$ Eq.~\ref{eq:notatri} actually defines a selection rule imposing integer quantum numbers $n$ over half integers (see Table~\ref{table} for some numerical values of $m_\pm$).  
\begin{table}[h]
\begin{tabular}{l||c|c|c|c|c||c}
  & $n=\frac{1}{2}$ & $n=1$ &  $n=\frac{3}{2}$ & $n=2$ & $n=\frac{5}{2}$ & $n=10$\\
\hline
\hline
$m_+$& 34.75 & 138 & 309.75 & 550 & 858.75 & 13710 \\
\hline
$m_-$& 33.75 & 136 & 306.75 & 546 & 853.75 & 13690 \\
\end{tabular}
\caption{Table giving the first quantum numbers  $m_\pm$ as a function of the azimuthal number $n$ for   $\alpha^{-1}=137$ and $b=1$. It can easily be proven that  half integer values $n=1/2, 3/2...$ fail to give integer values for $m_\pm$ and thus contradict Eq.~\ref{eq:notatri}. Observe that $m_\pm$ grow  rapidly  with $n$.  }
\label{table}
\end{table}
\\
\indent Of course, we experimentally know that $\alpha^{-1}$ is not an integer. Writing $\alpha^{-1}=137+ \xi$ we have   $\xi\simeq 3.5999$ $10^{-2}$. It can be easily checked that with this value of $\alpha$ we still fulfill Eq.~\ref{eq:notatri} by slightly modifying  the value of  $b\simeq 1$.    More precisely, writing  Eq.~\ref{eq:notatri} as 
 \begin{eqnarray}
  m_\pm= \frac{b \tilde{n}^2}{\alpha} \pm \tilde{n}.
	\label{eq:notaquad}
\end{eqnarray}    with $\tilde{n}=\frac{m_+-m_-}{2}\in \mathbb{N}$ we see that the condition for $b/\alpha$ to be an integer $q$ reads $b= \frac{q}{137+\xi}\simeq \frac{q}{137}(1-\frac{\xi}{137})$.   If we take $q=137$ we get $b\simeq 1-\frac{\xi}{137}\simeq 1- 3. 10^{-4}$. This  value is indeed very close from $b=1$ and therefore in this model $n=b\tilde{n}$ must be an integer up to a fluctuation $\delta n/\tilde{n}= \frac{\xi}{137}$ in order to satisfy Eq.~\ref{eq:notaquad}. This defines a small deviation with respect to the Bohr and Sommerfeld quantization postulate and therefore shows the limitation of our model. Finally,  we point out that the case  $b\simeq 1$ is only the simpler choice. However, similar conclusions concerning Eq.~\ref{eq:notaquad} are obtained if we consider the integers  $q=b/\alpha=2,3...$. In particular,   if $1/\alpha\in \mathbb{N}$ and  $b\in \mathbb{N}$ we again obtain a selection rule forcing the use of integer values for $n$.   

%%%%%%%%%%%%%%%%%%
\subsection{Solutions for the $u-$field}\label{sec:3c}
In order to solve the wave equation for $u$ in a central potential, we first write it as $u=u_++u_-$ with the two eigenmodes 
\begin{equation}
	u_\pm(t,\vb{x}) = \phi_\pm(\vb{x})\mathrm{e}^{-\mathrm{i}\omega_\pm t}.
\end{equation} solutions of Eq.~\ref{eq:field_motion}. The wave equation for a eigenstate reads: 
\begin{equation}
	\bqty{\pqty{\omega_\pm + \frac{\beta}{r}}^2 + \grad^2}\phi_\pm(\vb{x}) = 0.
\end{equation} where we defined the modified fine structure constant $\beta=\xi\frac{e^2}{4\pi}=\xi \alpha$ taking into account the fact that the charge $e'=\xi e$ associated with the $u-$ field is not neccessarily identical to $e$ (i.e., $\xi\neq 1$). 
In spherical coordinates $r,\theta, \varphi$, we seek separable eigenmodes having the general form:
\begin{equation}
	\phi_\pm(\vb{x}) = A_\pm R_{l_\pm}(r)Y_{l_\pm,\pm m_\pm}(\theta, \varphi).
\end{equation} where $Y_{l,m}(\theta, \varphi)$ is a spherical harmonic function with integer quantum number $l,m$ (i.e., $|m|\leq l$): 
\begin{equation}
	Y_{l,m}(\theta,\varphi) := P_{l}^m(\cos\theta)\mathrm{e}^{\mathrm{i}m\varphi}.
\end{equation}
with $P_{l}^m(\cos\theta)$ an associated Legendre polynomial (the irrelevant normalization constant has been here absorbed in the $A_\pm$ constant). 
For the present problem the radial function $R_{l_\pm}(r)$ follows the equation
\begin{equation}
	\bqty{\frac{1}{r^2}\partial_r(r^2\partial_r) - \frac{l_\pm(l_\pm + 1) - \beta^2}{r^2} + \frac{2\beta\omega_\pm}{r} + \omega_\pm^2}R_{l_\pm}(r) = 0.
	\label{eq:lapl_R}
\end{equation}
Using the substitution
\begin{align}
	l_\pm(l_\pm + 1) - \beta^2 &= l_\pm^\prime(l_\pm^\prime + 1)
	% \beta\omega &= m^\prime\beta^\prime\\
	% \omega^2 &= 2m^\prime\omega^\prime
\end{align}
we get
\begin{equation}
	l_\pm^\prime = -\half + \sqrt{\pqty{l_\pm+\half}^2 - \beta^2}
\end{equation} (which reduces to $l_\pm$ if  $\beta=0$)
and equation \ref{eq:lapl_R} becomes
\begin{equation}
	\bqty{\frac{1}{r^2}\partial_r(r^2\partial_r) - \frac{l_\pm^\prime(l_\pm^\prime + 1)}{r^2} + \frac{2\beta\omega_\pm}{r} + \omega_\pm^2}R_{l_\pm^\prime}(r) = 0.
\end{equation}
% The asymptotical solution for large values of $r$ is 
% \begin{equation}
% 	F_{l}(r) \approx \sqrt{\frac{2}{\pi}}\frac{1}{r}\sin\pqty{k r + \delta_{l^\prime}(r) - \frac{\pi}{2}l^\prime}
% \end{equation}
% with
% \begin{equation}
% 	\delta_{l^\prime}(r) = \frac{1}{k}\frac{q^\prime m^\prime}{\hbar^2}\ln(2kr) + \arg\bqty{\Gamma\pqty{l^\prime + 1 - \frac{\mathrm{i}}{k}\frac{q^\prime m^\prime}{\hbar^2}}}
% \end{equation}.
The solution for the radial function reads
\begin{equation}
	R_{l_\pm^\prime}(r) = e^{\mathrm{i}\omega_\pm r}r^{l_\pm^\prime}M(l_\pm^\prime + 1 - \mathrm{i}\beta, 2l_\pm^\prime + 2, -2\mathrm{i}\omega_\pm r)
\end{equation}
where $M(a,b,z):={}_1F_1(a,b,z)$ is the  Kummer confluent hypergeometric function which is a regular solution of $z\frac{d^2M}{d z^2}+ (b-z)\frac{dM}{d z}- a=0$.\\ 
\indent The asymptotic solution for large values of $r$ is 
\begin{equation}
		R_{l_\pm^\prime}(r) \approx C_{l_\pm^\prime}\frac{\sin(\omega_\pm r - \frac{\pi}{2}l_\pm^\prime + \delta_{l_\pm^\prime})}{\omega_\pm r}\label{assy}
\end{equation}
with $\delta_{l_\pm^\prime} = \beta\ln(2\omega_\pm r) + \eta_{l_\pm^\prime}$, $\eta_{l_\pm^\prime}=\arg(\Gamma(l_\pm^\prime + 1 - \mathrm{i}\beta))$
and $C_{l^\prime}$ a normalization constant reading:
\begin{eqnarray}
C_{l_\pm^\prime}=\frac{e^{i\eta_{l_\pm^\prime}}e^{-\beta \pi/2}}{(2\omega_\pm)^{l_\pm^\prime}}\frac{\Gamma(2l_\pm^\prime+2)}{\Gamma(l_\pm^\prime+1-\mathrm{i}\beta)}.
\end{eqnarray}
\indent Having obtained the eigensolutions for our wave equation, we return to the case of the transparency regime. As we explained, we need a superposition of two counter-propagating modes $u_\pm$ in order to reproduce the total field of Eq.~\ref{eq:field_e_cos}.
Since our two modes can be written as
\begin{equation}
	u_\pm(t,\vb{r}) = A_\pm R_{l^\prime_\pm}(r)P_{l_\pm}^{\pm m_\pm}(\cos\theta)\mathrm{e}^{\mathrm{i}\pqty{\pm m_\pm \varphi - \omega_\pm t}}\label{equat}
\end{equation}
with $A_\pm$ two normalization constants and $m_\pm\geq 0$. We have $P_{l_-}^{- m_-}(\cos\theta)=(-1)^{m_-}\frac{(l_--m_-)!}{(l_-+m_-)!}P_{l_-}^{m_-}(\cos\theta)$. We also have $P_l^m(0)=(-1)^{\frac{l+m}{2}}\frac{(l+m-1)!!}{(l-m)!!}$ if $l+m$ is even and $P_l^m(0)=0$ if $l+m$ is odd. It follows for a specific orbit in the equatorial plane (with radius $r_n$ and a polar angle $\theta = \pi/2$) that in order to recover Eq.~\ref{eq:field_e_cos} we must impose:
\begin{equation}
	A_\pm R_{l^\prime_\pm}\qty(r_{n})P_{l_\pm}^{\pm m_\pm}(0) = \frac{u_0}{2}.
\end{equation} 
These two constants $A_\pm$, which are defined for the specific radius $r_n$, will also give us the link between the amplitudes of the field $u$ and the oscillatory motion $z$ of the particle through the constraint:
\begin{equation}
	z_0=u_0.
\end{equation}
Having done that and using Eq.~\ref{eq:w_vec_freqbis}, the final field on the particle’s orbit is 
% \begin{equation}
% 	u\qty(t,r_n,\frac{\pi}{2},\varphi) = 2A\mathrm{e}^{\mathrm{i}\pqty{n\varphi - \omega_n t}}\cos\bqty{\pqty{\omega_n - \frac{\alpha}{r_n}}\pqty{r_n\varphi - v_n t}}	
% \end{equation}
\begin{equation}
	\begin{split}
		u\qty(t,r_n,\frac{\pi}{2},\varphi) &= z_0\mathrm{e}^{\mathrm{i}\pqty{\frac{n}{b}\varphi - \frac{N-\alpha}{br_n} t}}\\
		&\times\cos\bqty{\frac{N}{b}\pqty{r_n\varphi-\frac{n}{N}t} }
	\end{split}	\label{eq:field_e_cosnew}
\end{equation} which recovers Eq.~\ref{eq:field_e_cos}.\\
\indent Some comments must be done concerning the $u-$field solutions obtained so far.\\ 
\indent First, observe that the form of the equation for the $u-$modes was not very constraining.  Indeed, the value of the charge $e'$ is not specified. In particular, one can consider the chargeless fluid $e'=0$ and the solution  $R_l(r)$ reads now:
\begin{eqnarray}
	R_{l_\pm}(r) = e^{\mathrm{i}\omega_\pm r}r^{l_\pm}M(l_\pm + 1, 2l_\pm + 2, -2\mathrm{i}\omega_\pm r)\nonumber\\
	=\frac{1}{(2\omega_\pm)^{l_\pm}}\frac{(2l_\pm +1)!}{l_\pm !}j_{l_\pm}(\omega_\pm r)
\end{eqnarray}  where $j_l(x)=(-x)^l(\frac{1}{x}\frac{d}{dx})^l(\frac{\sin{x}}{x})$ are well known spherical Bessel functions.  For large value of $r$ we have 
\begin{equation}
R_{l_\pm}(r)\simeq \frac{1}{(2\omega_\pm)^{l_\pm}}\frac{(2l_\pm +1)!}{l_\pm !}\frac{\sin(\omega_\pm r - \frac{\pi}{2}l)}{\omega_\pm r},
\end{equation} which agrees with Eq.~\ref{assy} with $l_\pm^\prime=l_\pm$, $\eta_{l_\pm}=\delta_{l_\pm}=0$.\\
%%%%%%%%%%%%%%%%
\begin{figure}[h]
\begin{center}
\includegraphics[width=8.5cm]{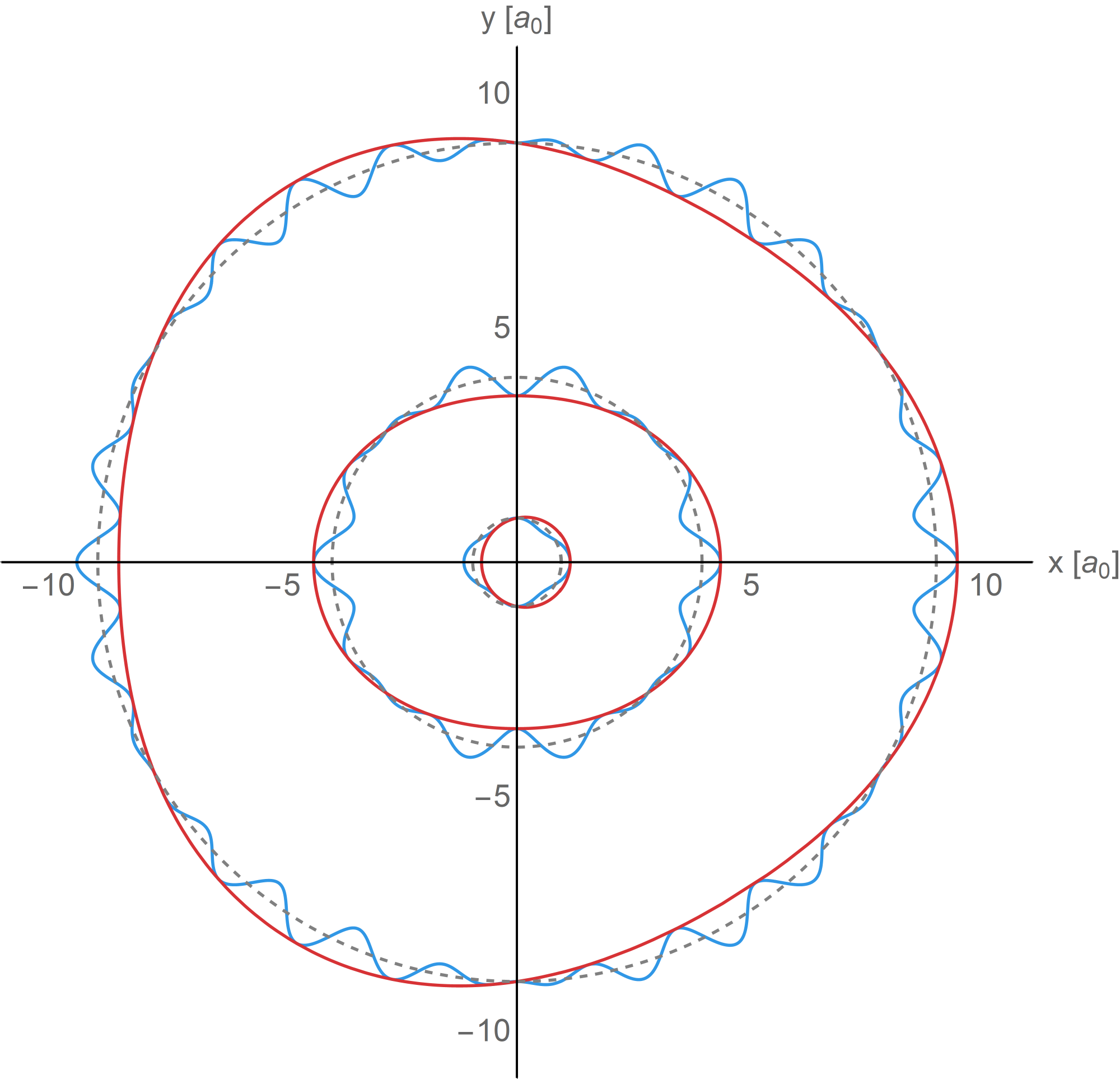} 
\caption{Parametric representation of the $u-$wave along the particle trajectory in the $x-y$ equatorial plane.  The particle trajectory for $n=1,2,3$ are the dashed (gray color) circles with constant radius $r_n$ given by Eq.~\ref{eq:p_radius}. The $u-$wave for $n=1,2,3$  are represented (i.e., blue curves) as the parametric curves $[x_n(\varphi)=R_n(\varphi)\cos{\varphi},y_n(\varphi)=R_n(\varphi)\sin{\varphi} ]$ with $R_n(\varphi)= r_n+\Delta  \Re[u(t=0,r=r_n,,\theta=\frac{\pi}{2},\varphi)] $ and where $\Delta$ is a constant used for graphical convenience. For comparison, the de Broglie phase guiding wave at the given time $t=0$ are similarly  shown (i.e., red curves) as parametric curves with radius $R_n(\varphi)= r_n+\Delta z_0\cos{(n\varphi)}$ where $z_0$ is here chosen real (see Eq.~\ref{eq:field_e_cosnew}). In this figure we imposed $b=1$, $\alpha=\beta=1/3$, $\omega_0=0$ and we used the maximal values of the quantum numbers $l_\pm=m_\pm$ (see also Figs.~\ref{figure2}-5).  } \label{figure1}
\end{center}
\end{figure}
%%%%%%%%%%%%%%%%
\indent Alternatively, and as briefly alluded to in Sec.~\ref{sec:2b}, instead of Eq.~\ref{eq:field_motion} for the $u-$field, one could consider the continuous spectrum of the Klein-Gordon equation.  This leads to the wave equation
\begin{equation}
	\bqty{\pqty{\omega_\pm + \frac{\beta}{r}}^2 + \grad^2}\phi_\pm(\vb{x}) = \omega_0^2\phi_\pm(\vb{x}).\label{Klein}
\end{equation} For $\omega_\pm\geq \omega_0$ we obtain a solution like  Eq.~\ref{equat} but where the radial function reads now 
\begin{equation}
	R_{l_\pm^\prime}(r) = e^{\mathrm{i}\tilde{\omega}_\pm r}r^{l_\pm^\prime}M(l_\pm^\prime + 1 - \mathrm{i}\tilde{\beta}_\pm, 2l_\pm^\prime + 2, -2\mathrm{i}\tilde{\omega}_\pm r)
\end{equation}
with $\tilde{\omega}_\pm=\sqrt{\omega_\pm^2-\omega_0^2}$ and $\tilde{\beta}_\pm=\beta\omega_\pm/\tilde{\omega}_\pm$. This constitutes the most general form of our wave field associated with a continuous spectrum for $\omega_\pm$. Remark, that one could also consider the discrete spectrum for $\omega_\pm\leq \omega_0$. However, discrete eigenmodes for the wave equation appear too restrictive for the present model.  In particular, it becomes in general impossible to find a simple `phase harmony' condition between the energy spectrum $E_n$ of the particle and the wave spectrum   $\omega=\frac{\omega_++\omega_-}{2}$ unless for very particular and contrived cases. Therefore, we avoided the use of discrete eigenmodes. \\
%%%%%%%%%%%%%%%%
\begin{figure}[h]
\begin{center}
\includegraphics[width=8.5cm]{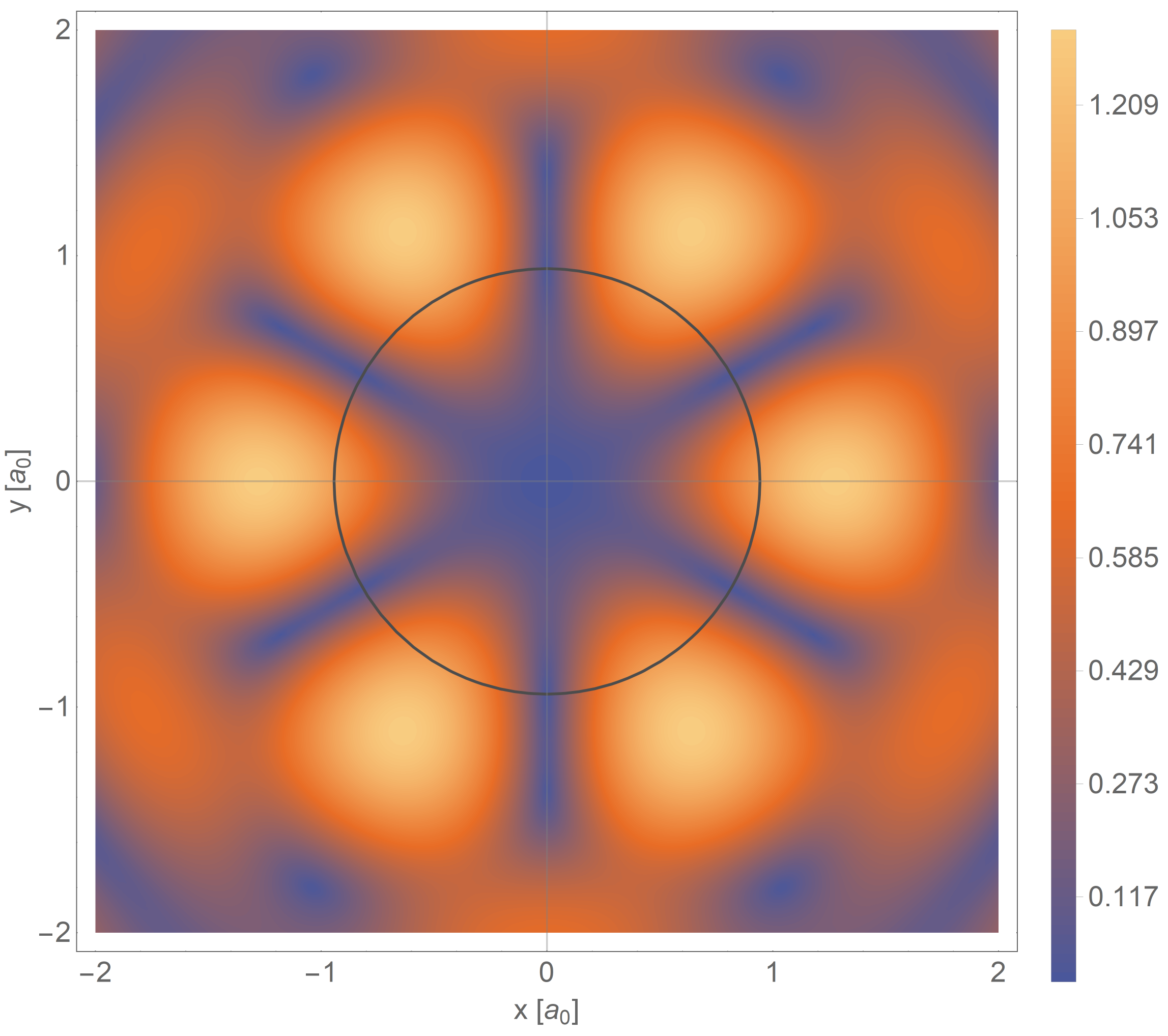} 
\caption{Intensity  map of the $u-$field in the $x-y$  equatorial plane for the quantum number $n=1$ (ground state).  The arbitrary normalized field intensity is defined as $I_n(x,y)=|u(t=0,r,\theta=\frac{\pi}{2},\varphi)|^2$ with $[x=r\cos{\varphi},y=r\sin{\varphi} ]$. The parameters and conditions for calculating the field are the same as for Fig.\ref{figure1}.  The black circle represents the particle trajectory and the spatial dimensions are normalized to the Bohr radius value $a_0$.   } \label{figure2}
\end{center}
\end{figure}
%%%%%%%%%%%%%%%%
\indent An other related comment about the $u-$modes concerns the Legendre associated polynomials $P_{l_\pm}^{\pm m_\pm}(\cos\theta)$ that are involved in Eq.~\ref{equat}. Indeed, the mathematical structure of the wave doesn't constrain very much the choice of the $l_\pm$ values despite the fact that we must have $m_\pm\leq l_\pm$.    %%%%%%%%%%%%%
\begin{figure}[h]
\begin{center}
\includegraphics[width=8.5cm]{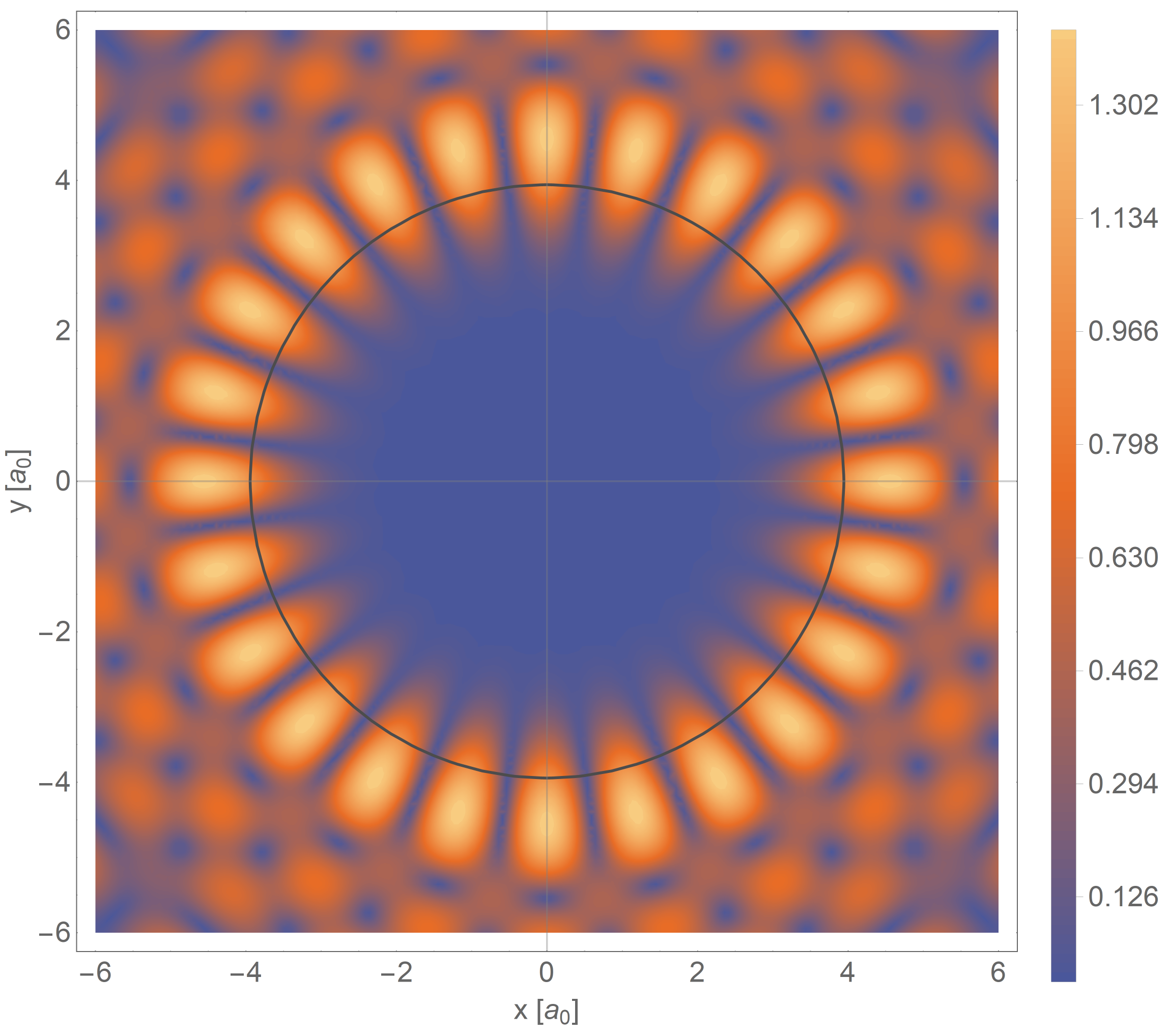} 
\caption{Same as for Fig.\ref{figure2} but with the quantum number $n=2$.} \label{figure3}
\end{center}
\end{figure}
%%%%%%%
This once again shows that many solutions for the $u-$field sare compatible with the present atomic model. Perhaps the most natural or intuitive choice in the context of the Bohr-Sommerfeld theory would be to impose   $l_+=m_+$ and $l_-=-m_-$ leading to $P_{m_+}^{m_+}(\cos\theta)=(-1)^{m_+}(2m_+-1)!!\sin^{m_+}\theta$ and $P_{m_-}^{-m_-}(\cos\theta)=\frac{(2m_--1)!!}{(2m_-)!}\sin^{m_-}\theta$. Indeed, with this choice the $u-$wave is strongly confined in the equatorial plane $\theta=\pi/2$ containing the orbit for $l_\pm\gg 1$.  We actually believe or hope that   in the high quantum number limit the equatorial plane acts as a dynamical attractor for the particle trajectory in the non-transparent regime $\mathcal{N}\neq 0$.  In other words, due to the force $-\mathcal{N}^\ast\boldsymbol{\nabla}u+ cc.$ acting upon the particle, the strong field  gradient in the spatial region $\theta \simeq \pi/2$ is expected to attract the particle in the equatorial plane.   Further work on stability is necessary to confirm or not this hypothesis.\\              
%%%%%%%%%%%%%%%%%%%%%%%%%%%%%%%%
\begin{figure}[h]
\begin{center}
\includegraphics[width=8.5cm]{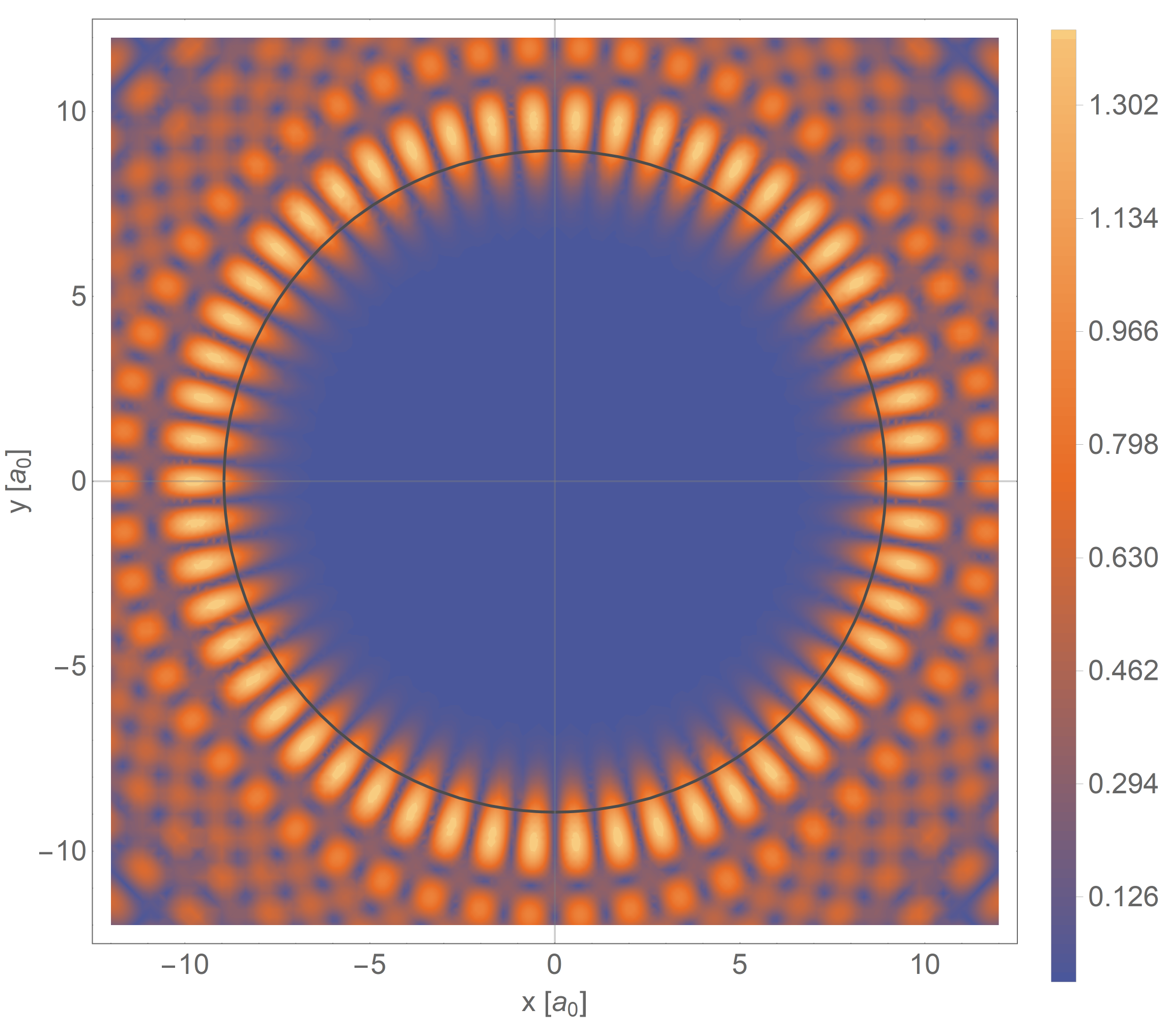} 
\caption{Same as for Fig.\ref{figure2} but with the quantum number $n=3$.} \label{figure4}
\end{center}
\end{figure}
%%%%%%%%%%%%%%%%$$
%%%%%%%%%%%%%%%%
\begin{figure}[h]
\begin{center}
\includegraphics[width=8.5cm]{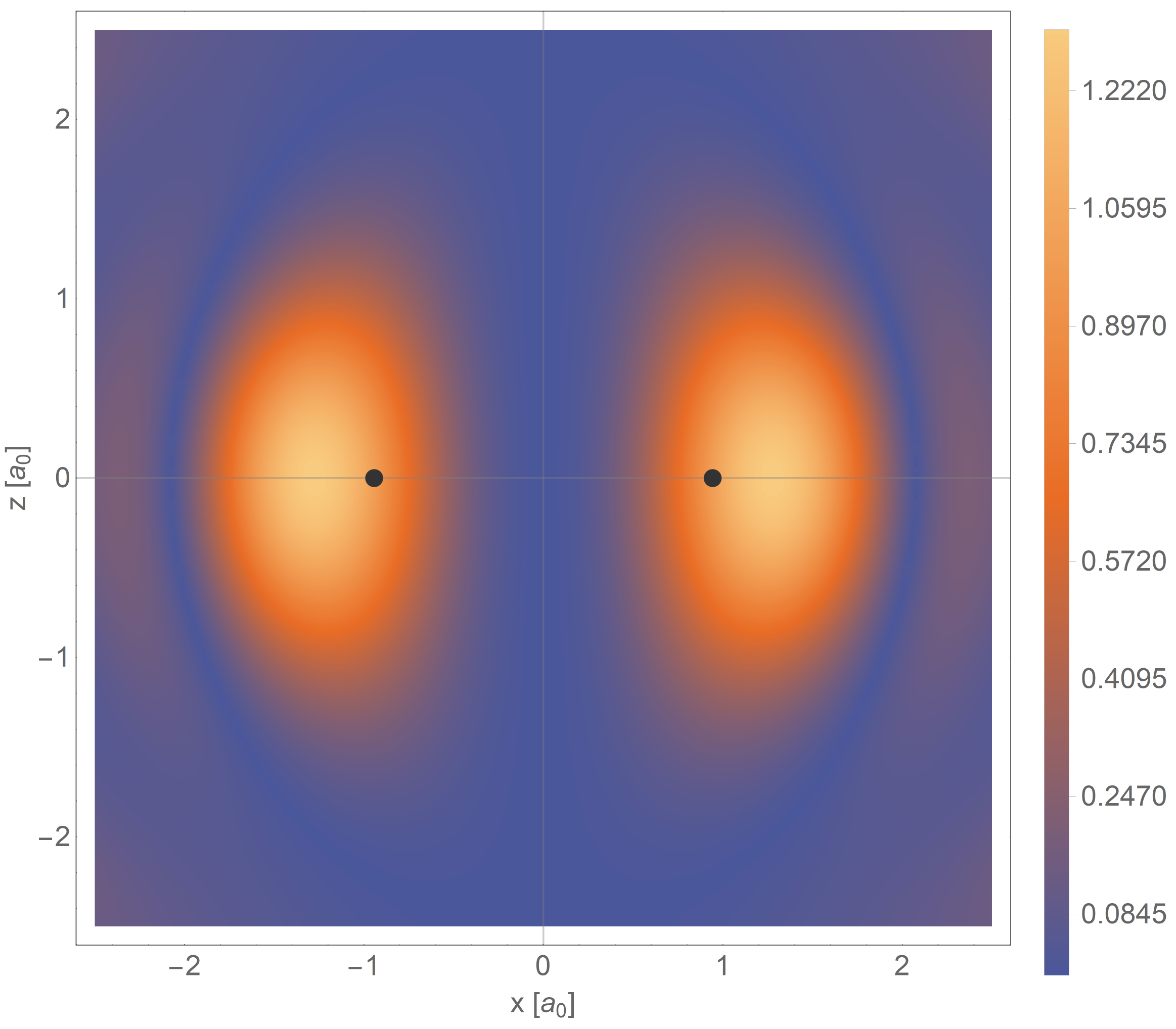} 
\caption{Intensity  map of the $u-$field in the $x-z$  transverse plane for the quantum number $n=1$ (ground state). The parameters are the same as for the other figures. The two black dots are the intersection of the circular orbit of radius $r_1$.} \label{figure5}
\end{center}
\end{figure}
%%%%%%%%%%%%%%%%
\indent  To illustrate the complete dynamics we first show in Fig.~\ref{figure1}  at the given time $t=0$ a parametric representation of the propagative $u-$field along the particle trajectory for the three first energy levels  $n=1,2$ and 3.  We compare the total field (blue curves) with the phase field (red curves) used in the paradigmatic de Broglie atomic model.  In particular,  we see that the total field involves  fast oscillations with shorter wavelengths that are  associated with the group-wave, i.e., envelope wave, propagation (see Eq.~\ref{eq:field_e_cosnew}). This situation is reminiscent of the analysis already obtained in our previous article for a 1D string mechanical analog \cite{Drezet2020}.  Furthermore,  the longer wavelength modulations in the $u-$field (red curves) are  associated with the faster-than-light phase-wave of de Broglie in full agreement with the 1D string model \cite{Drezet2020}.  For the illustrations we used the particular conditions  $b=1$, $\alpha=\beta=1/3$, $\omega_0=0$ leading to easy observation of the field modulations. Furthermore,  we used the maximal values of the quantum numbers $l_\pm=m_\pm$. As explained, this choice is motivated by the semiclassical approximation which in quantum mechanics is described by the Brillouin-Wentzel-Kramers (BWK) theory working for high quantum numbers.  Graphically, the $u-$field intensity is actually strongly confined near the particle trajectory as shown in Figs.~\ref{figure2}, \ref{figure3}, \ref{figure4} in the $x-y$ equatorial plane   for   $n=1,2,$ and 3 respectively.   The same effect occurs in the $x-z$ plane  as shown in Fig.\ref{figure5} for the case   $n=1$  with a doughnut shape for the intensity profile. This  analysis shows that already for small quantum numbers   the $u-$field is strongly confined near the Bohr-Sommerfeld  trajectory.   This feature is of course an interesting specificity of our model. More studies are needed to further understand the implications of these results for discussing  more complicated atomic motions.    
%%%%%%%%%%%%
\section{Perspectives and conclusions}\label{sec:4}
\indent   The model proposed in this article is directly motivated by the first `phase-harmony' wave-mechanics proposed by de Broglie between 1923-1925~\cite{debroglie1923,debroglie1925}. 
Both models are dualistic in nature coupling a point-like particle to an extended guiding-wave.  The key idea of the phase harmony hypothesis is the synchronization between the local clock associated with the particle and the wave-field computed at the position of the particle. In our model, this condition is summarized by the holonomic constraint Eq.~\ref{eq:constr} $z(\tau)=u(\pcl{x}(\tau))$ locking the phase and amplitude of the $u-$wave  with those of the internal clock oscillation $z(\tau)$.\\
\indent We emphasize that for de Broglie in his early work the physical meaning of the phase-wave was not very clear. The specificity of our model is the introduction of a $u-$field having a physical content like the classical electromagnetic field or the gravitational metric  in general relativity. This is clearly reminiscent of the double-solution program developed by de Broglie~\cite{debroglie1927,debroglie1956}. More precisely, in the double solution of  de Broglie one postulates the existence of a physical  $u-$field guiding the particle considered as a localized `accident' in the wave (i.e., a singularity\cite{Drezet2021}). Our model is more specific than the one proposed by de Broglie.   First,  there is indeed a phase-wave coming from the sum of two counterpropagating modes $u_\pm$ in Eq.~\ref{eq:field_e_cos}. However,  we have also a group wave (the cosine term  in Eq.~\ref{eq:field_e_cos})  and both are essential for guiding  the particle. Indeed, the subluminal group wave guides the particle since the constancy of its amplitude during the particle motion allows us to fulfill one part of the holonomic constraint: $|z(\tau)|=|u(\pcl{x}(\tau))|$. The superluminal phase-wave also guides the particle since it fixes the  phase-harmony condition: $\arg{[z(\tau)]}=\arg{[u(\pcl{x}(\tau))]}$, and it allows us to fix the dynamics of the particle in order to recover the Bohr-Sommerfeld quantization formula $\oint P_ndx=2\pi n$. This separation of the $u-$wave into a phase and group contributions is thus fundamental in our approach since  it explains why we must consider two quantum numbers $n$ and $N\gg n$ (see Eq.~\ref{eq:nota}) instead of only one quantum number $n$ in the old de Broglie theory~\cite{debroglie1923,debroglie1925}.\\ 
\indent Remarkably, our `entangled' wave/particle dynamics, with these two quantum numbers, leads to strong constraints on the particle properties like  the effective `dressed' mass $m_{\mathrm{eff.}}$ (i.e., Eq.~\ref{mass}), the internal oscillator amplitude $\vqty{z_0}$ (i.e., Eq.\ref{amplitude})  which depend on the number $n$, and the electric charge $e=\sqrt{(4\pi\alpha)}$ which depends on both $n$ and $N$ through the fine structure constant $\alpha = \frac{n^2}{N}$  (see Eq.~\ref{eq:charge}). As we showed in Sec.\ref{sec:3b}  the constraint imposed on the particle charge  $e$ allows us to define a selection rule for the quantum number $n$ in the regime   $b=1$.    With this choice we get an interesting relation  (i.e., Eq.~\ref{eq:notaquad}) prohibiting half integer quantum numbers $n=1/2,3/2...$ in the Bohr-Sommerfeld formula $\oint P_ndx=2\pi n$.
Moreover, we stress once more that the value of the $b$ constant used in our model is not imposed by the theory itself but must be better seen as an initial or boundary condition for the whole coupled system particle-wave.\\
\indent We also stress that the present theory requires us to find two eingensolutions $u_\pm(t,\vb{x}) = \phi_\pm(\vb{x})\mathrm{e}^{-\mathrm{i}\omega_\pm t}$ obeying a Klein-Gordon wave equation $D^2 u(x) +\omega_0^2 u_\pm(x)=0$, i.e., Eq.~\ref{Klein}, where $D_\mu=\partial_\mu+ie'A_\mu$ depends on a field charge $e'=\xi e$ that is in general different from the particle charge $e$. The theory is however not constraining very much the choice of the parameters $\omega_0$ and $e'$ in  Eq.~\ref{Klein}.  Our main requirement is to be able to find a combination    
$\omega=\frac{\omega_++\omega_-}{2}$ such that we can recover the Bohr-Sommerfeld quantization spectrum $E_n$ through the equality $E_n=b\omega$.  Moreover, since we are considering the continuous spectrum $\omega_\pm\geq \omega_0$ we get from the definition $E_n\geq b\omega_0$ and thus  with Eq.~\ref{eq:p_energy}:
\begin{equation}
E_n=m_{\mathrm{eff.}}\sqrt{1 - \frac{\alpha^2}{n^2}}\geq m_{\mathrm{eff.}}\sqrt{1 - \alpha^2}\geq b\omega_0.
	\label{eq:p_energybis}
\end{equation}
It is thus always possible to find a particle mass $m_{\mathrm{p}}$ in the effective mass $m_{\mathrm{p}}(1 + \sigma\Omega_{\mathrm{p}}^2\vqty{z_0}^2)$ in order to fulfill this condition.  This is the case in particular if $\omega_0=0$ which is also the simpler choice.\\
\indent It is interesting to watch the problem of the form of the $u-$wave equation from a different perspective.    Indeed,  let us  write the field $u_\pm$ in polar coordinates
\begin{equation}
	u_\pm(t,\vb{r}) = f_\pm(t,\vb{r})\mathrm{e}^{\mathrm{i}\Phi_\pm(t,\vb{r})}
\end{equation} (with $f_\pm$ and $\Phi_\pm$ real)
which lets us separate the Klein-Gordon equation \ref{eq:field_motionKG} into two parts:
\begin{equation}
	\partial^\mu\bqty{f_\pm^2(\partial_\mu \Phi_\pm + e' A_\mu)} =0
\end{equation}
and
\begin{equation}
	(\partial \Phi_\pm + e' A)^2 = \omega_0^2+\frac{\square f_\pm}{f_\pm}.\label{HamiltonJacobiQ}
\end{equation}
The second equation is reminiscent of the quantum version of the Hamilton-Jacobi equation introduced by de Broglie in his double solution and pilot-wave mechanics~\cite{debroglie1927,Valentini,debroglie1956}. The term $Q_\pm=\frac{\square f_\pm}{f_\pm}$ is called quantum potential and characterizes the difference between the quantum Eq.~\ref{HamiltonJacobiQ} and the classical equation $(\partial \Phi_\pm + e' A)^2 = \omega_0^2$ or in other words the difference between wave mechanics and the Eikonal equation of geometrical optics.\\ 
\indent In the present case, with $\omega_\pm=-\partial_t \Phi_\pm$ and  $k_\pm=\pm\frac{1}{r\sin \theta}\partial_\phi \Phi_\pm=\frac{m_\pm}{r\sin \theta}$  we obtain
 \begin{equation}
	(\omega_\pm + \frac{\beta}{r})^2-k_\pm^2 = \omega_0^2+Q_\pm.\label{HamiltonJacobiQB}
\end{equation}Moreover, from Eq.~\ref{hypo} we have along the particle orbit $r=r_n,\theta=\pi/2$ the condition 
$k_\pm = \omega_\pm + \frac{\alpha}{br_n}=\frac{m_\pm}{r_n}$, and by comparing with Eq.~\ref{HamiltonJacobiQB} we obtain 
\begin{equation}
Q_\pm(t,r_n,\phi,\theta=\pi/2)=(\frac{m_\pm}{r_n})^2-(\frac{m_\pm+\beta-\frac{\alpha}{b}}{r_n})^2	-\omega_0^2.\label{HamiltonJacobiQC}
\end{equation}
Remarkably,   if we impose $\omega_0=0$ and    $b\beta=\alpha$ (i.e., $be'=e$) we obtain rigorously $Q_\pm(t,r_n,\phi,\theta=\pi/2)=\frac{\square f_\pm}{f_\pm}=-\frac{\boldsymbol{\nabla}^2 f_\pm}{f_\pm}=0$ along the particle trajectory. In other words   the quantum Hamilton-Jacobi Eq.~\ref{HamiltonJacobiQ} reduces to the classical one along the orbit.  We believe   this is another motivation for the case $be'=e$, and $\omega_0=0$. Together with the condition $b=1$ this implies the wave equation 
\begin{equation}
(\partial +ieA)^2 u(x)=0
\label{final1}
\end{equation} where $e'=e$.\\
\indent A last point that we want to briefly comment concerning our model is about causality.  Indeed,  in order to work   our model involves  two   waves $u_\pm$ specially tuned in order to    reproduce the phase matching  condition of de Broglie and thereby the Bohr-Sommerfeld quantization formula.  Moreover,   this specific field $u=u_++u_-$ is in some sense conspiratorial or better `superdeterministic'. This issue about superdeterminism has recently been  the subject of many interesting discussions in the  context of Bell inequality and quantum nonlocality~\cite{Palmer2020,Vervoort2012}.   It is therefore not unreasonable to further study this possibility in order to develop a more sophisticated quantum model  using a $u-$field. This idea will be developed  in a subsequent work in preparation.\\  
\indent To summarize our work, we developed a model for a $u-$wave (i.e., solution of Eq.~\ref{final1}) guiding a particle.  The guiding dynamics is reminiscent of the old   phase-wave introduced by de Broglie  in order to justify the Bohr-Sommerfeld quantization condition  $\oint P_ndx=2\pi n$. More precisely Our model leads to a physical justification of this guidance condition by introducing a holonomic constraint $u(x(\tau))=z(\tau)$ between the wave and an internal degree of freedom $z(\tau)$ of the particle.  In turn,  the $u-$wave factorizes into a phase and group wave guiding the particle.  In particular, if $b=1$, the phase wave 
\begin{equation}
\Psi(t,\mathbf{x})\sim e^{i(P_n x-E_n)t}=e^{iS_n(t,\mathbf{x})}\label{trucmuch}
\end{equation}
is similar to the semiclassical solution of the Klein-Gordon equation            
 \begin{equation}
(\partial +ieA)^2 \Psi(x)=-m_{\mathrm{eff.}}^2\Psi(x)
\label{final2}
\end{equation} where the effective particle mass $m_{\mathrm{eff.}}$ is included. If $b\neq 1$ the correspondence is not direct between the phase wave reading generally  
$\Psi(t,\mathbf{x})\sim e^{iS_n(t,\mathbf{x})/b}$ and the de Broglie/Schr\"odinger wave since we have now $(\partial +ieA)^2 \Psi(x)=-\frac{m_{\mathrm{eff.}}^2}{b^2}\Psi(x)$ which involves a kind of effective Planck constant $b$. Moreover, even in this case it is possible to recover quantum mechanics if we now define the quantum de Broglie/Schr\"odinger wave as $\Psi_{\textrm{quantum}}(t,\mathbf{x}):=\Psi^b(t,\mathbf{x})\sim e^{iS_n(t,\mathbf{x})}$ which satisfies Eq.~\ref{final2}.\\
\indent More generally, the duality between $u(x)$ solution of Eq.~\ref{final1} and $\Psi(x)$ solution of Eq.~\ref{final2} is clearly reminiscent of de Broglie's double solution~\cite{debroglie1927,debroglie1956}. In the end, the particle dynamics we obtain is identical to the one predicted by the de Broglie-Bohm pilot-wave  theory applied to the Klein-Gordon equation~\cite{debroglie1927,Valentini,Bohm1952,Hiley} in the semiclassical regime.  This pilot-wave theory predicts that the velocity of the particle guided by the $\Psi-$wave solution of Eq.~\ref{final2} is given by 
\begin{equation}
\pcl{\mathbf{v}}(t)=-\frac{\boldsymbol{\nabla}S(t,\pcl{\mathbf{x}}(t))-e\mathbf{A}(t,\pcl{\mathbf{x}}(t))}{\partial_t S(t,\pcl{\mathbf{x}}(t))+eV(t,\pcl{\mathbf{x}}(t))}\label{Bohmian}
\end{equation} where $S(t,\pcl{\mathbf{x}}(t))$ is the phase of  $\Psi(t,\pcl{\mathbf{x}}(t))$ computed at the position $\pcl{\mathbf{x}}(t)$ of the particle at time $t$.  In order to recover our model with the pilot-wave theory we must put $\mathbf{A}=0$ and $-eV=\alpha/r$ and we consider only a semiclassical solution of Eq.~\ref{final2} leading to the Bohr-Sommerfeld formula  $\oint P_ndx=2\pi n$ where Eq.~\ref{Bohmian} reduces to Eq.~\ref{eq:p_velocityC}.\\ 
\indent Moreover, contrary to usual quantum mechanics (and also pilot-wave theory),  our model for the $u-$wave is valid for any integer quantum  numbers $n=1,2...$ and not only for large integers $n\gg 1$ required in the BWK semiclassical approximation of  Eq.~\ref{final2}.  This shows the limitation of our mechanical analogy of quantum mechanics.\\ 
\indent A different interesting feature of our model is the quantization of the constant $\alpha^{-1}$ which is required in order to satisfy the set of coupled equations.   This property is remarkable since it shows that coupling a guiding wave to a particle in order to reproduce quantum mechanics can lead to strong constraints on the physical parameters.  This is to be expected since the $u-$wave needs interferences and a resonance condition in order to reproduce Bohr's quantization formula.  This feature was ignored in the original phase-wave model of de Broglie where a mechanical description  like the one proposed in Sec.~\ref{sec:2b} was missing. Here, we have two quantum numbers $m_\pm$ rather than one as it was in the first proposal of de Broglie.  This is due to the fact that we need two waves to reproduce the guidance formula of de Broglie. In the end,  this constraint on $\alpha$ has strong physical consequences.   If a model like ours has to be taken seriously it apparently implies a strong fine-tuning on the parameters used, which might be related to the freedom we have regarding the parameter $b$ in our model.   We dont have here an explanation for this fact but it suggests a cosmological explanation perhaps related to some (weak) anthropic principle.     \\    
\indent  At the same time, the model for the $u-$wave demonstrates that it is in principle possible to reproduce some important features of quantum mechanics with a classical and deterministic analogy. Furthermore, unlike in the conventional Copenhagen interpretation where the very notion of an orbit is ill defined,  here the particle path is deterministic and continuous in the four-dimensional space-time. We believe this result to be in the direct continuation of early works by de Broglie~\cite{debroglie1927,debroglie1956}  and more recent ones on hydrodynamic mechanical analogs by Couder and Bush~\cite{Couder2006,Bush2015,Bush2015b,Bush2021}.\\
\indent In this context, our theory offers some interesting potentialities in order to develop realistic mechanical demonstrator  for the Bohr atomic  model.  As we explained  in Sec.~\ref{sec:2b} our model is a direct generalization of our previous article \cite{Drezet2020} where a 1D model is used for a transverse wave propagating  along an elastic string and guiding a particle.   In Sec.~\ref{sec:2b} we pointed out that a 2D mechanical analog using a vibrating membrane coupled to a particle could  constitute a realistic demonstrator of our model. We believe that other physical analogies could be developed along this direction.  For example,  we could imagine an hydrodynamic analog  with a particle coupled to acoustic waves propagating in a  spherical or torroidal tank.  Optical traps and tweezers using  laser beams are also good candidates. Optical vortices with well defined angular orbital momenta can be nowadays easily generated (e.g., \cite{Ashkin,DrezetVortex}) at least in 2D. Since our model can easily be developped in 2D space this suggests interesting experimental developments. In particular, optical traps in liquids have the potentiality, i.e., coupled  to small Brownian particles in water, to create stabilized circular orbital motions of particles  fulfilling the Bohr-Sommerfeld quantum condition. We believe that all these interesting issues deserve further analysis.\\   
\indent To conclude it is interesting to go back to de Broglie's double solution research program:  In the 1950's  de Broglie   returned to his double solution after 25 years. In his new version of the theory \cite{debroglie1956} he wrote the $u-$field as
\begin{eqnarray}
u(x)=u_0(x)+v(x) \label{ansatz}
\end{eqnarray}  
where $u_0(x)$ was a strongly singular wave associated with the particle and $v(x)$ a base wave guiding   the point-like singularity and solution of a linear wave equation such as the Klein-Gordon or Schr\"odinger equation. For de Broglie   this base wave is proportional to  the usual quantum   $\Psi-$wave of quantum mechanics:   $v(x)=C\Psi(x)$  (with $C$ a constant). Our model shows strong similarities with this idea since our $u-$field indeed guides the particle (acting as a kind of singularity).  Moreover, our dynamics is satisfying  the action-reaction principle  since there is a coupling between the wave and the particle such that if the transparency regime $\mathcal{N}=0$ is not satisfied a new $u-$field solution of Eq.~\ref{eq:field_sourceKG}, i.e., $D^2 u(x) =-\frac{1}{T}\int d\tau \mathcal{N}(\tau)\delta^4(x - \pcl{x})$ will be emitted by the particle and this in turn will modify the motion of the point-like singularity. In the transparency regime considered in this article  the wave and the particle peacefully ignore each other in order to satisfy a guidance condition which is reminiscent of the pilot-wave interpretation (at least in the semi-classical regime). We believe that the knowledge of the source field  emitted if $\mathcal{N}\neq 0$ could play a role in order to describe optical transitions between the different energy levels of the atom. This requires to include radiation damping due to electromagnetic self-interaction of the moving electron and clearly opens interesting possibilities for future extensions of the present model.\\

%%%%%%%%%%%%%%%%%%%%%%                             
\appendix
\section{Derivation of the equations of motion}
We first introduce an affine parameter $\lambda$ along the particle trajectory such that $\sqrt{\pqty{x^\prime}^2}\dd{\lambda} = \dd{\tau}$ (in the following $f^\prime$ denotes a derivative $\frac{d}{d\lambda}f(\lambda)$) and write the action \ref{eq:action} of the whole system:  
\begin{equation}
	\begin{split}
		I = &-\int\bqty{m_{\mathrm{p}} - \frac{m_{\mathrm{p}}\sigma}{2}\pqty{\frac{\vqty{z^\prime(\lambda)}^2}{(x^\prime)^2} - \Omega_{\mathrm{p}}^2 \vqty{z(\lambda)}^2}}\sqrt{\pqty{x^\prime}^2}\dd{\lambda} \\
		&+ \int \left\lbrace \mathcal{N}(\lambda)\bqty{z(\lambda) - u(x_{\mathrm{p}}(\lambda))}^*\right. \\
		&+ \left. \mathcal{N}^*(\lambda)\bqty{z(\lambda) - u(x_{\mathrm{p}}(\lambda))}\right\rbrace\sqrt{\pqty{x^\prime}^2}\dd{\lambda} \\ 
		&- e\int A(x_{\mathrm{p}}(\lambda))x^\prime_{\mathrm{p}}(\lambda)\dd\lambda + T\int(Du)(Du)^*\dd^4 x.
	\end{split}	
\end{equation} In order to write the Euler-Lagrange equations for the `particle' variables $y(\lambda):=[z(\lambda)$, $z^\ast(\lambda)$, $\mathcal{N}(\lambda)$, $\mathcal{N}^\ast(\lambda)$ and $x(\lambda)]$ we write the previous action integral as $I=\int d\lambda \mathcal{L}(y,y^\prime)+ T\int(Du)(Du)^*\dd^4 x$ where the four volume integral $T\int(Du)(Du)^*\dd^4 x$ is actually irrelevant since it is independent of $y$ and $y^\prime$. The Lagrangian function $\mathcal{L}(y,y^\prime)$  reads:
\begin{equation}
	\begin{split}
		\mathcal{L}(y,y^\prime)= &-\bqty{m_{\mathrm{p}} - \frac{m_{\mathrm{p}}\sigma}{2}\pqty{\frac{\vqty{z^\prime(\lambda)}^2}{(x^\prime)^2} - \Omega_{\mathrm{p}}^2 \vqty{z(\lambda)}^2}}\sqrt{\pqty{x^\prime}^2}\\
		&+  \left\lbrace \mathcal{N}(\lambda)\bqty{z(\lambda) - u(x_{\mathrm{p}}(\lambda))}^*\right. \\
		&+ \left. \mathcal{N}^*(\lambda)\bqty{z(\lambda) - u(x_{\mathrm{p}}(\lambda))}\right\rbrace\sqrt{\pqty{x^\prime}^2}\\ 
		&- eA(x_{\mathrm{p}}(\lambda))x^\prime_{\mathrm{p}}(\lambda)
	\end{split}	
\end{equation}
This allows us to write Euler-Lagrange equations $\frac{d}{d\lambda}\frac{\partial L}{\partial y^\prime}=\frac{\partial L}{\partial y}$:\\ 
--i) For $\mathcal{N}$ we obtain:
\begin{equation}
	\pdv{\mathcal{L}}{\mathcal{N}^*} =\sqrt{\pqty{x^\prime}^2}\pqty{ z(\lambda) - u(x_{\mathrm{p}}(\lambda)) }= 0,
\end{equation} which leads to the holonic condition $z(\lambda) - u(x_{\mathrm{p}}(\lambda)=0$ (a similar equation is obtained for the complex conjugate variables).\\

--ii)For $z$ we have:
\begin{equation}
	\pdv{\mathcal{L}}{z^*} = \bqty{-\frac{m_{\mathrm{p}}\sigma}{2}\Omega_{\mathrm{p}}^2 z(\lambda) + \mathcal{N}(\lambda)}\sqrt{\pqty{x^\prime}^2}
\end{equation}
\begin{equation}
	\pdv{\mathcal{L}}{z^{\prime *}} = \frac{m_{\mathrm{p}}\sigma}{2}\frac{z^\prime(\lambda)}{\sqrt{\pqty{x^\prime}^2}}
\end{equation}
\begin{equation}
	\frac{m_{\mathrm{p}}\sigma}{2}\bqty{\dv{\lambda}\pqty{\frac{z^\prime(\lambda)}{\sqrt{\pqty{x^\prime}^2}}} + \Omega_p^2 z(\lambda)\sqrt{\pqty{x^\prime}^2}} = \mathcal{N}(\lambda)\sqrt{\pqty{x^\prime}^2}
\end{equation}
In particular if we choose $\lambda=\tau$ where $\tau$ is the proper time we have  $\sqrt{\pqty{x^\prime}^2}=1$ and we obtain 
\begin{equation}
	m_{\mathrm{p}}\frac{\sigma}{2}\pqty{\frac{d^2}{d\tau^2}z(\tau)+ \Omega_p^2 z(\tau)}= \mathcal{N}(\tau).
\end{equation} Again, we stress that similar equations are easily obtained  for the complex conjugate variables $z^\ast$, $N^\ast$.\\

--iii) Similarly, for   $x_p$ (which is the only real valued variable in the dynamics) we get: 
\begin{equation}
	\pdv{\mathcal{L}}{x^\mu} = -(\mathcal{N}\partial_\mu u^\ast+\mathcal{N}^\ast\partial_\mu u) \sqrt{\pqty{x^\prime}^2} -e\partial_\mu A_\nu {x^\prime}^\nu,
\end{equation}
\begin{eqnarray}
	\pdv{\mathcal{L}}{{x^\prime}^\mu} = -m_{\mathrm{p}}\left( 1 + \frac{\sigma}{2}\pqty{\frac{\vqty{z^\prime(\lambda)}^2}{(x^\prime)^2} + \Omega_{\mathrm{p}}^2 \vqty{z(\lambda)}^2}\right)\frac{{x^\prime}_\mu}{\sqrt{\pqty{x^\prime}^2}}\nonumber\\
	 -\Big(eA_\mu +\mathcal{N}^\ast(z-u)+\mathcal{N}(z^\ast-u^\ast)\Big)\frac{{x^\prime}_\mu}{\sqrt{\pqty{x^\prime}^2}}.\nonumber\\
\end{eqnarray}Moreover after introducing the holonomic conditions $z(\lambda)=u(x(\lambda))$, $z^\ast(\lambda)=u^\ast (x(\lambda))$ we deduce:
\begin{eqnarray}
	\frac{m_{\mathrm{p}}}{\sqrt{\pqty{x^\prime}^2}}\frac{d}{d\lambda}\pqty{\left( 1 + \frac{\sigma}{2}\pqty{\frac{\vqty{z^\prime(\lambda)}^2}{(x^\prime)^2} + \Omega_{\mathrm{p}}^2 \vqty{z(\lambda)}^2}\right)\frac{{x^\prime}_\mu}{\sqrt{\pqty{x^\prime}^2}}}\nonumber\\
	=\mathcal{N}^\ast\partial_\mu u+\mathcal{N}\partial_\mu u^\ast+eF_{\mu\nu}\frac{{x^\prime}^\nu }{\sqrt{\pqty{x^\prime}^2}}. \nonumber\\
\end{eqnarray}
In particular if $\lambda =\tau$ we obtain:
\begin{eqnarray}
	m_{\mathrm{p}}\frac{d}{d\tau}\pqty{\pqty{1 + \frac{\sigma}{2}\pqty{\vqty{\frac{d}{d\tau}z(\tau)}^2+ \Omega_{\mathrm{p}}^2 \vqty{z(\tau)}^2}}\frac{d}{d\tau}x_\mu(\tau)}\nonumber\\
	=\mathcal{N}^\ast\partial_\mu u+\mathcal{N}\partial_\mu u^\ast+eF_{\mu\nu}\frac{d}{d\tau}x^\nu . \nonumber\\
\end{eqnarray}

--iV) For the field variables $u(x)$ and $u(x)^\ast$ we rewrite the action integral as: $I=\int d^4x \mathfrak{L}(u,\partial u , u^\ast, \partial u^\ast,x)+ ...$
 where the dots are irrelevant terms independent of $u,\partial u , u^\ast, \partial u^\ast$ variables, and $\mathfrak{L}(u,\partial u , u^\ast, \partial u^\ast,x)$ is the Lagrangian density:  
 \begin{eqnarray}
 \mathfrak{L}(u,\partial u , u^\ast, \partial u^\ast,x)=T(Du)(Du)^\ast \nonumber\\
+ \int_{(C)} \Big( \mathcal{N}(\lambda)\bqty{z(\lambda) - u(x_{\mathrm{p}}(\lambda))}^* \nonumber\\
+  \mathcal{N}^*(\lambda)\bqty{z(\lambda) - u(x_{\mathrm{p}}(\lambda))}\Big)\sqrt{\pqty{x^\prime}^2}\delta^4\pqty{x - x_p(\lambda)}\dd{\lambda}\nonumber \\ 	
\end{eqnarray}
  where an integral along the trajectory $C$ of the particle has been included for convenience (it represents an explicit $x$ dependence in $ \mathfrak{L}$). We deduce
\begin{equation}
	\begin{split}
		\frac{1}{T}\pdv{\mathfrak{L}}{u^*} = &-\int_{(C)}d\lambda \mathcal{N}(\lambda)\sqrt{\pqty{x^\prime}^2}\delta^4\pqty{x - x_p(\lambda)} \\
		&- \mathrm{i} e A_\mu\pqty{\partial^\mu + \mathrm{i} e A^\mu}u
	\end{split}
\end{equation}
and
\begin{equation}
	\frac{1}{T}\pdv{\mathfrak{L}}{\partial_\mu u^*} = \pqty{\partial^\mu + \mathrm{i} e A^\mu}u
\end{equation}
\begin{equation}
	\frac{1}{T}\partial_\mu \pdv{\mathfrak{L}}{\partial_\mu u^*} = \partial_\mu\partial^\mu u + \mathrm{i} e\partial_\mu\pqty{A^\mu u}
\end{equation} which leads to the Euler-Lagrange equation:
\begin{equation}
	\begin{split}
		0 &= \partial_\mu\pdv{\mathfrak{L}}{\partial_\mu u^*} - \pdv{\mathfrak{L}}{u^*} \\
		&= T\bqty{\partial_\mu\partial^\mu u + \mathrm{i} e\partial_\mu\pqty{A^\mu u} + \mathrm{i} e A_\mu\pqty{\partial^\mu + \mathrm{i} e A^\mu}u}\\
		&+ \int_{(C)}d\lambda \mathcal{N}(\lambda)\sqrt{\pqty{x^\prime}^2}\delta^4\pqty{x - x_p(\lambda)}
	\end{split}
\end{equation}
and in the end we get
\begin{equation}
	D_\mu D^\mu u = -\int_{(C)}d\lambda \frac{\mathcal{N}(\lambda)}{T}\sqrt{\pqty{x^\prime}^2}\delta^4\pqty{x - x_p(\lambda)}
\end{equation}
The line integral alond $C$ can be written in a simpler form if $\lambda=\tau$ and we obtain 
\begin{equation}
	D_\mu D^\mu u = -\int_{(C)}d\tau \frac{\mathcal{N}(\tau)}{T}\delta^4\pqty{x - x_p(\tau)}.
\end{equation} 	Alternatively, we can use $\lambda=t'$ where $t'$ is a laboratory time for the particle and  we then obtain
 \begin{eqnarray}
	D_\mu D^\mu u(t,\mathbf{x}) = -\int_{(C)}dt' \frac{\mathcal{N}(t')}{T}\sqrt{1-v^2(t')}\delta(t-t')\nonumber\\ \times\delta^3\pqty{\mathbf{x} - \mathbf{x}_p(t')}
	=-\frac{\mathcal{N}(t)}{T}\sqrt{1-v^2(t)}\delta^3\pqty{\mathbf{x} - \mathbf{x}_p(t)}.\nonumber\\ 
\end{eqnarray}
\indent We stress that we can directly introduce a covariant form of the Euler-Lagrange equations and get the same wave equation, by taking $Du$ and $D^*u^*$ as our independent variables instead of $\partial u$ and $\partial u^*$~\cite{Lewis}.\\
%%%%%%%%%%%%%%%%%%
\section*{Data Availability Statement}

Data sharing is not applicable to this article as no new data were created or analyzed in this study.
%%%%%%%%%%%%%%ù
\bibliographystyle{eplbib} 

\end{document}